\documentclass[a4paper,11pt]{article}
\usepackage{geometry}
\usepackage{makecell}
\usepackage{amsmath, amssymb, amsthm, bbm, bm}
\usepackage{booktabs}
\usepackage{placeins}
\usepackage{natbib}
\usepackage{graphicx}
\usepackage{url}
\usepackage{authblk}
\usepackage{float}
\usepackage{multirow}
\usepackage{caption}

\title{Gaussian Process Priors with Markov Properties for Effective Reproduction Number Inference}

\author[1]{Jessalyn N. Sebastian}
\author[1]{Volodymyr M. Minin}

\affil[1]{Department of Statistics, University of California, Irvine}

\date{}

\begin{document}

\maketitle

\abstract{Many quantities characterising infectious disease outbreaks --- like the effective reproduction number ($R_t$), defined as the average number of secondary infections a newly infected individual will cause over the course of their infection --- need to be modelled as time-varying parameters. It is common practice to use Gaussian random walks as priors for estimating such functions in Bayesian analyses of pathogen surveillance data. In this setting, however, the random walk prior may be too permissive, as it fails to capture prior scientific knowledge about the estimand and results in high posterior variance. We propose several Gaussian Markov process priors for $R_t$ inference, including the Integrated Brownian Motion (IBM), which can be represented as a Markov process when augmented with its corresponding Brownian Motion component, and is therefore computationally efficient and simple to implement and tune. We use simulated outbreak data to compare the performance of these proposed priors with the Gaussian random walk prior and another state-of-the-art Gaussian process prior based on an approximation to a Mat\'ern covariance function. We find that IBM can match or exceed the performance of other priors, and we show that it produces epidemiologically reasonable and precise results when applied to county-level SARS-CoV-2 data.}

\section{Introduction}
Careful real-time monitoring of infectious disease outbreaks is key to enabling policymakers to implement effective mitigation strategies \citep{cori_key_2017, niu_collaboration_2021, van_kerkhove_epidemic_2012}. While surveillance data like time series of cases, hospitalisations, or deaths are revealing in of themselves, mathematical modelling that incorporates multiple data streams and available prior knowledge provides deeper insights into the current dynamics of the outbreak's spread. In this paper, we propose and contrast several smoothing models for doing inference on time-varying transmission, and we assess their implications for retrospective and real-time epidemic monitoring.

Measuring whether an outbreak is growing or subsiding can be of particular interest to public health officials when assessing the current intensity of the disease's spread, as well as the impact of interventions like masking and school closures. The effective reproduction number, $R_t$, is one such measure. The effective reproduction number is the expected number of individuals a newly infected individual will subsequently infect; when $R_t$ is less than 1, it indicates that the outbreak is diminishing, and when it exceeds 1, the outbreak is worsening. This makes $R_t$ a straightforward, interpretable indicator of the current state of an epidemic. As a consequence, many methods for estimating $R_t$ from surveillance data like cases and diagnostic tests already exist and are being used in public health response around the world \citep{Cori_Kucharski_2024}.

Most of these methods operate within a fully Bayesian framework. Regardless of the structure of the likelihood connecting the observed data to the underlying model, all of these methods require defining a prior distribution over possible $R_t$ trajectories. Since $R_t$ is inherently a function of time, these priors effectively determine the temporal correlation structure of the function. A common approach is to use a first-order Gaussian random walk (RW1) \citep{conway_joint_2024, Parag_2021, improving_rt, jin_combining_2024, epinow2, epinowcast, epidemia}, a discretised and scaled Wiener process, in which each successive $R_t$ (typically with some link function due to nonnegativity constraints) is modelled as a Gaussian distribution with mean at the previous $R_t$ and a variance proportional to the amount of time elapsed between observations. The RW1 is an appealing prior because it is flexible, easy to implement, and has interpretable hyperparameters.

In the setting of $R_t$ inference, the data involved can be particularly noisy. Although the RW1 is flexible, it may be too permissive for $R_t$ inference. There is plenty of prior knowledge about the structure of $R_t$, so using the RW1 may be leaving information that could sharpen inference on the table; models using the RW1 prior can suffer from wider-than-necessary Bayesian credible intervals (BCIs) due to the roughness of its sample paths.

These issues have prompted some modellers to look elsewhere for appropriate $R_t$ priors. Gaussian process priors commonplace in Bayesian spatio-temporal statistics and machine learning, like those with a squared exponential or Matérn kernel, can be used to impose smoothness. Bayesian computation with these functional priors, however, is significantly more costly than when using the simple RW1. Numerous approximations have been developed to make Gaussian process priors more tractable, like reduced-rank approximations using the Nyström method or local approximations with model averaging like mixture of experts (see, for example, \cite{liu_when_2020} and \cite{Rasmussen_Williams_2008} for an overview). Popular epidemic modelling package EpiNow2 \citep{epinow2} employs a Hilbert space approximation for stationary covariance kernels (HSGP), as developed by \citet{Solin_Sarkka_2019} and extended for easier application in oft-used modelling software like Stan by \citet{Riutort-Mayol_etal_2022}. A potential downside of this method, and other Gaussian process approximation methods like it, is the need to choose settings for the approximation in addition to priors for the parameters of the Gaussian process. These settings can be difficult to interpret practically, and may therefore be challenging to practitioners without significant mathematical background.

The purpose of this study is to investigate potential priors that have the benefits of both types of priors discussed above. From the perspective of infectious disease monitoring, desirable priors for $R_t$ balance several considerations: computational tractability, epidemiological plausibility, interpretability, and ease of implementation. We turn to Gaussian process priors with Markov properties, or conditional independence properties, meaning that the value of the process at any point in time conditional on the rest of the time points only depends on its value at neighbouring points. This structure allows for the joint density to be expressed as a product of conditional transition densities between neighbouring points, which simplifies the evaluation of the probability density and circumvents the need for computationally expensive matrix inversions. We compare the Ornstein-Uhlenbeck (OU), Second-Order Gaussian Random Walk (RW2), and Integrated Brownian Motion (IBM) priors with the RW1 and the HSGP with a Matérn kernel, within the framework of a simple hierarchical Bayesian model that takes counts of cases as data. We consider both retrospective and real-time inference, comparing the distinct effects of smoothing assumptions on the entire inferred time-varying $R_t$ curve and on only the latest $R_t$ inference. In simulations, we find that the IBM and RW2 priors match or outperform the RW1 and HSGP in both accuracy and precision of retrospective $R_t$ posterior inference. The IBM prior additionally produces inference comparable to the HSGP and more precise than the RW1 in real time, and is more likely to correctly identify increasing transmission. We compare the results of models using these priors on California COVID-19 data collected at the county level, and we demonstrate that they produce reasonable results, but that posterior uncertainty around $R_t$ is higher with the RW1 than with smoother priors like the HSGP and IBM.

\section{Methods}

\subsection{Likelihood of Observed Cases} 

\label{sec:likelihood}

\begin{figure}[t]
    \centering\includegraphics[width = 0.8\textwidth]{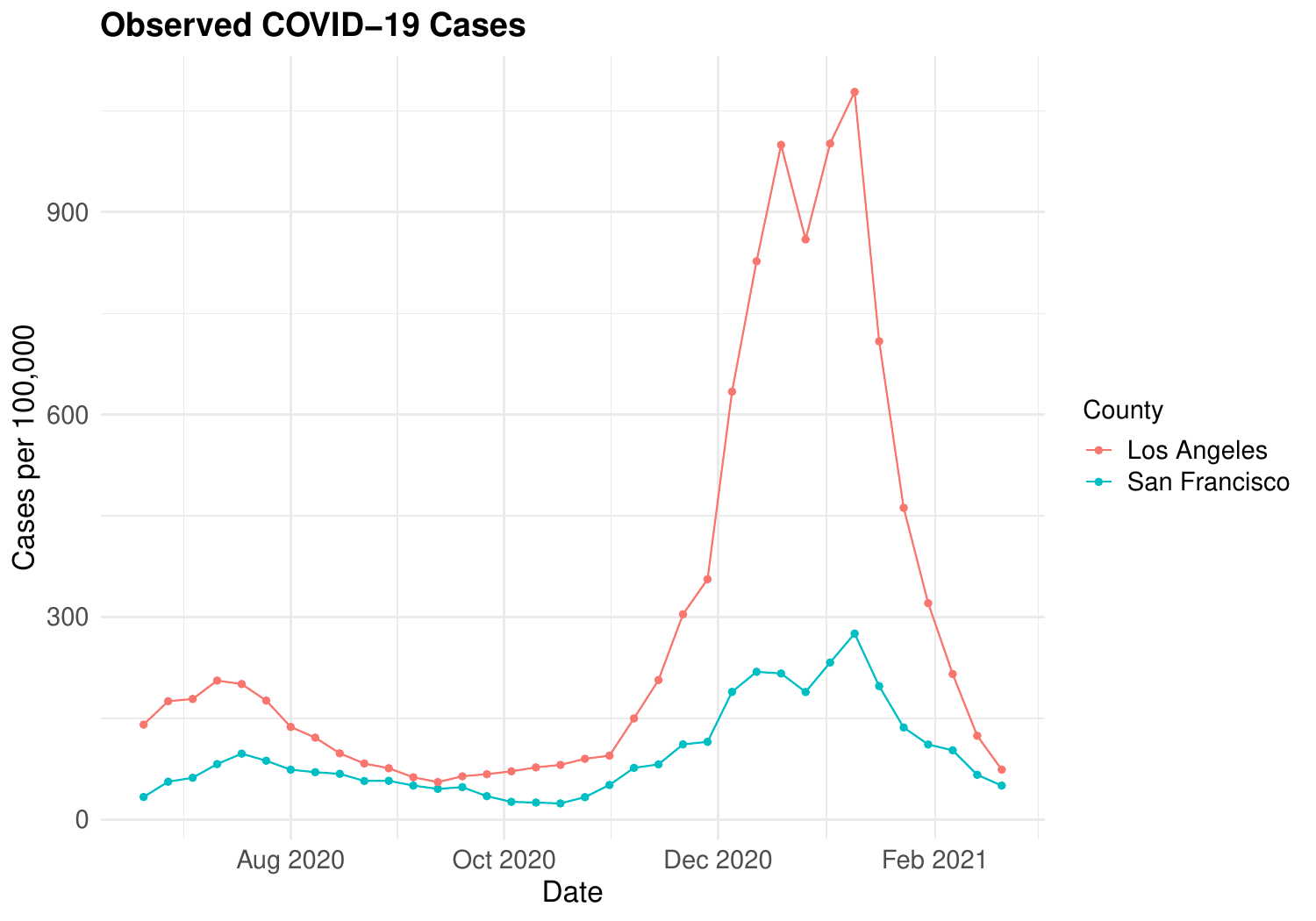}
    \caption{Time series of COVID-19 cases in the counties of San Francisco and Los Angeles, California, from June of 2020 through February of 2021.}
    \label{fig:cases_timeseries}
\end{figure}

The effective reproduction number is typically inferred using surveillance data like cases, hospitalisations, diagnostic tests, and/or deaths. The public health threat posed by the COVID-19 pandemic drove the production of types of datasets that have not historically been available, such as pathogen concentration in wastewater or in some cases digital contact tracing data, from which $R_t$ can also be estimated \citep{Cori_Kucharski_2024}. While these emerging data sources have expanded the scope of infectious disease modelling, traditional surveillance data remain popular for real-time outbreak monitoring, due primarily to their widespread public availability. The \citet{ca_open_data_portal}, for example, is a statewide data portal hosting data from many local government agencies, including COVID-19 surveillance data, like time series of observed cases, at the county level. Suppose we want to use these data to generate $R_t$ inference for several counties in California through Summer-Fall-Winter of 2020--2021 (Figure \ref{fig:cases_timeseries}). 

Speaking more generally, we assume that we want to infer $R_t$ for some pathogen over a discretely-indexed time frame $t = 1, ..., T$, and the data contain a time series of observed cases $\mathbf{O} = (O_1,O_2,..., O_T)$. Each element in $\mathbf{O}$ gives the number of newly observed cases of an infectious disease in the time interval given by the subscript.

The effective reproduction number governs changes in latent incidence, or the true, unobserved counts of infected individuals. Observed cases, in turn, should be modelled as some noisy realisation of a function of that latent incidence. This structure lends itself well to a hierarchical Bayesian model. 

Following \citet{epinow2}, \citet{epidemia}, and  \citet{rtlive}, among others, we use a negative binomial distribution to model observed cases $\mathbf{O}$ conditional on current and prior incidence ($I_t$ and $\mathbf{I}_{-n:t}$, respectively, where $n$ is the number of weeks of seeded incidence, discussed below). The use of the negative binomial likelihood is conventional, as it allows for modelling overdispersed counts. The expectation of observed cases is a function of true underlying incidence and the ascertainment rate $\rho$, or the fraction of latent incidence which is observed. The overdispersion of the negative binomial distribution at each time $t$ is given by $\kappa$. We then have

$$O_t\mid I_t, \mathbf{I}_{-n:t}, \rho, \kappa \sim \text{Neg-Binom}(\rho \times D_t, \kappa), \hspace{1em} t = 1,\dots, T,$$

\noindent with

$$D_t = \sum_{j=-n}^t I_j d_{t-j},$$

\noindent where $d_{t-j}$ are probabilities from a discretised delay distribution, and delays represent the time between infection of an individual and the subsequent detection of their case. In this study, we will treat delays as being the time between infection and testing, with no additional delay for reporting the results of the test. This is a reasonable modelling choice for the data we analyze in this paper, as it gives counts of cases by the date of their test, rather than the date of their ultimate reporting. We make the additional simplifying assumption that individuals test around the time they become infectious. This means that the $d_{t-j}$ are discretised probabilities from the latent period distribution, or the distribution of times between infection and becoming infectious. Then, the joint density for observed cases over the observational period takes the form

$$P(\mathbf{ O}\mid \mathbf{ I}, \rho,\kappa) = \prod_{t=1}^T P(O_t\mid I_t, \mathbf{I}_{-n:t}, \rho, \kappa),$$

\noindent since observed cases in each time interval are modelled as conditionally independent given true unobserved incidence. Parameter $\rho$ is given a log-normal prior based on a plausible range of proportions of true incidence that are observed, and $\kappa$ is given a truncated-normal prior. Poor choices of $\kappa$ result in poor Markov chain Monte Carlo (MCMC) convergence. This can be addressed in several ways; we follow \citet{improving_rt} in using a Bayesian thin plate regression spline to build a prior for $\kappa$ (see Appendix \ref{sec:kappa}). To be philosophically sound, the spline model should use data from a time and location that is expected to be similar to, but not the same as, the data to be modelled.

The model for unobserved incidence connects the time series of observed cases to the primary target of inference, the effective reproduction number. It is common to model seeded incidence, or incidence prior to the observational period, drawn from an exponential distribution \citep{epidemia}. For $t = -n, \dots, 0$, we draw $I_t$ from an exponential distribution with mean $\lambda$, which itself is given an exponential prior. During the observational period, that is, $t = 1,\dots, T$, we follow \citet{improving_rt} and model current incidence conditional on previous incidence and $R_t$ as a gamma distributed random variable:

$$I_t\mid \mathbf{I}_{-n:t}, R_t, \nu \sim \text{Gamma}\bigg(R_t\sum_{j = -n}^{t-1} g_{t-j}I_j\nu, \nu\bigg), \hspace{1em} t = 1,\dots, T,$$

\noindent where $g_t$ are probabilities from the discretised generation time distribution, or distribution of times between infection of an individual and the preceding infection of their infector, and $\nu$ represents the linear mean-variance relationship for the distribution of incidence. The mean of the gamma distribution comes from the well-known renewal equation \citep{fraser_estimating_2007}. The renewal equation is used in outbreak modelling to express the mode of propagation of a disease as a function of average transmissibility, infectious period length, and current incidence. In many cases, incidence is modelled deterministically with the renewal equation; we believe it is desirable to have some stochasticity in incidence, especially in small- and medium-sized populations \citep{improving_rt}. The joint model for incidence is then

$$P(\mathbf{ I}\mid \mathbf{ R},\nu,\lambda) = \bigg(\prod_{t=-n}^0 P(I_t\mid \lambda)\bigg)\bigg(\prod_{t=1}^T P(I_t\mid \mathbf{I}_{-n:t}, R_t, \nu)\bigg).$$

We have discussed our modelling choices for observed cases and latent incidence, and what remains is the prior for $R_t$, the primary subject of this paper. The model we have presented above is fairly generic, where most of our modelling choices are common, and their implications are well-understood in the field. We do not expect that minor changes to this likelihood structure will substantially change the relative impacts of each of the $R_t$ priors on the estimated posterior distributions.

\subsection{Priors for the Effective Reproduction Number}
\label{sec:rt_priors}

In this paper, we treat $R(t)$ as a continuous-time latent function, and we construct our priors based on the real-valued $\Gamma(t)=\log R(t)$, since the effective reproduction number is strictly nonnegative. We write $R_t$ for the value of $R(t)$ at a given observation time $t$, and similarly $\Gamma_t$ for the value of $\Gamma(t)$ at time $t$. The model works with the discretised vector:

$$\mathbf\Gamma = (\Gamma_{1}, \Gamma_{2},\dots,\Gamma_{T}).$$

A Gaussian process is a type of stochastic process where any finite collection of random variables drawn from the process follows a multivariate normal distribution. Continuous-time GP priors induce a multivariate normal distribution for $\mathbf\Gamma$ when evaluated on this grid, while discrete-time models specify the distribution of $\mathbf\Gamma$ directly.

A Gaussian Markov process is a specific case of a Gaussian process that additionally satisfies a Markov property. Conceptually, this property of conditional independence states that given the current state of the process, the future state is independent of prior states \citep[p. 13-17]{Rue_Held_2005}. The Markov property is particularly useful in simplifying computations, as it allows for computationally efficient modelling of dependencies over time or space. Using a general Gaussian process prior without a Markov property requires high-cost manipulations of the full covariance matrix of the discretised trajectory. When the prior has Markov structure, the joint density of the process can be computed more efficiently by simply multiplying conditional densities. It is because of their ease of implementation and computational benefits that we focus on Gaussian Markov processes in this paper. Below we introduce some specific priors and discuss them in further detail.

\subsubsection{Gaussian Random Walk (RW1)}

The Gaussian Random Walk is a time-homogeneous, first-order Gaussian Markov process, obtained by discretising and scaling a Wiener process, or Brownian Motion. This prior is given as

$$\Gamma_1 \sim \text{Normal}(\mu_1, \sigma^2_1),$$

$$\Gamma_t\mid \Gamma_{t-1} \sim \text{Normal}\bigg(\Gamma_{t-1}, \frac{\sigma^2}{T-1}\bigg), \; t=2,...,T,$$

\noindent where the choice of prior distribution of $\sigma$ corresponds to practical beliefs about the amount of fluctuation in $R_t$ over time. We have chosen to represent the conditional variance as $\sigma^2/(T-1)$ for ease of prior elicitation. If Var$(\Gamma_t\mid \Gamma_{t-1}) = \sigma^2/(T-1)$, then Var$(\Gamma_t\mid \Gamma_1) = \sum_{1}^{T-1} \sigma^2/(T-1) = \sigma^2$, by properties of Gaussian random walks. Then, the prior on $\sigma$ can be chosen based on the expected total possible fluctuation in $\Gamma_t$ over the observational period. The joint density of $\mathbf{\Gamma}$ for $t=1,\dots,T$ is then given by the product of conditional densities

$$P(\mathbf{\Gamma}\mid \mu_1, \sigma_1, \sigma) = P(\Gamma_1\mid \mu_1,\sigma_1)\prod_{t=2}^T P(\Gamma_t\mid \Gamma_{t-1}, \sigma).$$

The RW1 and its variants are common choices of $R_t$ prior. Some, like \citet{conway_joint_2024}, place the random walk directly on $R_t$. This makes it difficult to enforce nonnegativity; \citet{Parag_2021} does so by scaling the error by a factor proportional to the square root of the previous $R_t$. It is more typical to use some kind of link function. A log-link, as we use for all priors in this paper, is a common choice (\cite{improving_rt}, \cite{jin_combining_2024}). Similarly, modelling tools like EpiNow2 \citep{epinow2} and Epinowcast \citep{epinowcast} provide the option to use a random walk prior on $R_t$ with a log-link. A log-link implies no so-called carrying capacity; in other words, it is not bounded above. \citet{epidemia} view this as a downside and account for it using a scaled logit link instead.

\subsubsection{Ornstein-Uhlenbeck Process (OU)}

In practice, $R_t$ is generally known to drift toward and fluctuate around the critical value 1, especially for infectious diseases that become endemic \citep{Anderson_2016}. This prior knowledge can be reflected by the Ornstein-Uhlenbeck (OU) process, which is a modified Brownian Motion that approaches its long-term mean with drift proportional to distance from the mean.  The process $X_t^{OU}$ is defined by the stochastic differential equation

$$dX_t^{OU} = -\theta(X_t^{OU} - \mu)dt + \sigma B_t,$$

\noindent where $B_t$ is a Brownian motion, $\mu\in \mathbb{R}$ is the stationary mean, $\sigma$ scales the driving Brownian motion, and $\theta\in (0,\infty)$ controls the strength of reversion to the mean \citep[p. 74-77]{Gardiner_1985}. Brownian motion is then a special case of the OU process where $\theta = 0$. The OU process has a closed-form transition density given by

$$X_t^{OU} \mid X_0^{OU} \sim \text{Normal}\bigg(\mu + (X_0^{OU}-\mu)e^{-\theta t}, \frac{\sigma^2}{2\theta}(1-e^{-2\theta t}) \bigg) ,$$

 \noindent as stated in \citep[p. 74-77]{Gardiner_1985}. The OU process is temporally homogeneous, so the transition density above can be extended to any pair $X_{t+s}^{OU}\mid X_{s}^{OU}$.

 We let the prior for $\Gamma_t$ be an OU process on the log scale,  with long-term mean $\mu$ set equal to $0$ (thereby asserting that we expect $R_t$ to revert to $e^0=1$). The OU prior is then given by

$$\Gamma_1 \sim \text{Normal}(\mu_{1}, \sigma^2_{1}),$$

$$\Gamma_t\mid \Gamma_{t-1} \sim \text{Normal}\bigg(\Gamma_{t-1}e^{-\theta}, \frac{\sigma^2}{2\theta}(1-e^{-2\theta }) \bigg), \; t=2,...,T.$$

As with RW1, the joint density of $\mathbf{\Gamma}$ for $t=1,\dots,T$ can be given by the product of conditional densities

$$P(\mathbf{\Gamma}\mid \mu_1, \sigma_1, \sigma, \theta) = P(\Gamma_1\mid \mu_1,\sigma_1)\prod_{t=2}^T P(\Gamma_t\mid \Gamma_{t-1}, \sigma, \theta).$$

Hyperprior selection for the OU prior is more difficult than for the other Gaussian Markov processes presented here. The OU prior as given above takes two parameters: $\theta$, the strength of reversion to the mean, and $\sigma$, which scales the driving Brownian motion process. Since $\theta$ appears in both the mean and variance of the transition density, it is perhaps easiest to place a prior on $\theta$ based on the mean and then use the prior on $\sigma$ to scale the variance to reflect expected week-to-week (or day-to-day, depending on the timescale) variation in $\Gamma_t$.

The parameter $\theta$ is theoretically interpretable as the strength of the process's attraction to the limiting mean, but in practice it is not an easy task to give this parameter a concrete meaning. We consider $e^{-\theta}$, the term in the mean of the transition density that controls the change in mean compared to the prior $\log R_t$ value (which, in the RW1, would be taken as the mean itself). The prior on $\theta$ can then be constructed by proposing a range of plausible values for how much closer we would expect each successive $\Gamma_t$ to be to 0. Letting $\theta \sim \text{Exp}(1)$, for example, gives $e^{-\theta}$ a $\text{Uniform}(0,1)$ distribution. Then, using this, we can construct a prior on $\sigma$ based on conditional variance of $\Gamma_t$.

\subsubsection{Second-Order Gaussian Random Walk (RW2)}

Neither of the processes discussed above encourage smoothness of the sample paths. Both the RW1 and the OU processes are rough; their use as smoothing priors is rooted in the rescaling of time or conditional variance such that the discretely observed processes appear smooth. In general, we expect $R_t$ to vary smoothly with time, because it takes time for changes in susceptibility, mitigation measures, and emergence of new genetic variants to have an effect on transmission. This motivates the desire for a smooth, computationally inexpensive prior that maintains the ease of implementation and interpretability of the RW1 and OU priors.

A smooth prior with Markov structure can be obtained in discrete time by using a higher order model; that is, conditioning on multiple previous states rather than only one. Here we consider a Second-Order Gaussian Random Walk (RW2). While the idea behind the first-order RW1 is that increments, or successive differences, are zero-mean normal distributions, the RW2 gives increments of increments zero-mean normal distributions. This makes the process second-order Markov, as each successive state is dependent on the previous two; it is then a smoother process than the first-order RW1 since it holds more memory at each step.

RW2 models are interpretable and computationally inexpensive for observations at regular time intervals, but they are not appropriate for irregularly spaced data, which is not uncommon in infectious disease surveillance. To account for unevenly spaced data, some reweighting or other approximation can be done \citep{fahrmeir_dynamic_1997, lindgren_second-order_2008}. For the purposes of this paper, we consider only regularly spaced time series data. The RW2 prior for $\Gamma_t$ is then given by

$$\Gamma_1 \sim \text{Normal}(\mu_{1}, \sigma^2_{1}),$$
$$\Gamma_2 \mid \Gamma_1 \sim \text{Normal}(\Gamma_1, \sigma^2),$$
$$\Gamma_t\mid \Gamma_{t-1}, \Gamma_{t-2} \sim \text{Normal}(2\Gamma_{t-1} - \Gamma_{t-2}, \sigma^2), \; t=2,...,T,$$

\noindent\citep[p. 133-140]{Rue_Held_2005}. The joint density of $\mathbf{\Gamma}$ under this second-order Markov prior becomes

$$P(\mathbf{\Gamma}\mid \mu_1, \sigma_1, \sigma) = P(\Gamma_1\mid \mu_1, \sigma_1)P(\Gamma_2\mid\Gamma_1, \sigma)\prod_{t=3}^T P(\Gamma_t\mid\Gamma_{t-1}, \Gamma_{t-2}, \sigma).$$

\subsubsection{Integrated Brownian Motion (IBM)}

Integrated Brownian Motion (IBM), also known as the integrated Wiener process, is the time integral of Brownian motion:

$$X_t^{IBM}=\int_0^t B_sds,$$

\noindent where $X_t^{IBM}$ represents the IBM process and $B_t$ represents its corresponding Brownian motion \citep[p. 134]{Rue_Held_2005}. Since IBM is by definition the integral of a non-differentiable stochastic process, its sample paths are once differentiable, so this prior encourages smoothness like the RW2. IBM is, in fact, the continuous-time analogue of the RW2. This can be seen intuitively by considering the relationship between derivatives and differences: the time derivative of IBM is a Wiener process, while differencing a RW2 gives a RW1, which is a Wiener process in discrete time. Since IBM, unlike RW2, is defined in continuous time, the extension to irregularly spaced observations is natural.

Though IBM itself is not a Markov process, when augmented with the corresponding Brownian motion, the two can be modelled together as a two-dimensional process that is both Gaussian and Markov \citep{wecker_signal_1983}. This Markov representation of IBM lends itself well to being a smoothing prior, and its use can be motivated by the connection to smoothing splines as shown by \citet{wahba_improper_1978}. In spite of this, exact IBM has not been widely adopted as a smoothing prior; instead, some approximations are usually used (see, for example, \cite{lindgren_second-order_2008} and \cite{zhang_model-based_2024}). The transition density for $ \begin{bmatrix}
    B_t \\
    X_t^{IBM}
\end{bmatrix}$ is:

$$\begin{bmatrix}
    B_t \\
    X_t^{IBM}
\end{bmatrix} \bigg | \begin{bmatrix}
    B_s \\
    X_s^{IBM}
\end{bmatrix} \sim \text{Normal}_2 \Bigg( \begin{bmatrix}
    B_s \\
    X_s^{IBM} + (t-s) B_s
\end{bmatrix}, \begin{bmatrix}
    t-s & \frac{1}{2}(t-s)^2 \\
    \frac{1}{2}(t-s)^2 & \frac{1}{3}(t-s)^3
\end{bmatrix} \Bigg)$$

\noindent for $s\leq t$. We once again place the IBM prior on $\log R_t$, or $\Gamma_t,$ rather than $R_t$ directly. Letting $\Gamma_t'$ be the time derivative of $\Gamma(t)$ at time $t$, the prior becomes

$$\begin{bmatrix}
    \Gamma_1' \\
    \Gamma_1
\end{bmatrix} \sim \text{Normal}_2\Bigg(\begin{bmatrix}
    \mu_{1}' \\
    \mu_{1}
\end{bmatrix}, \begin{bmatrix}
    \sigma_1^2  & \sigma_1^4/2\\
   \sigma_1^4/2 &  \sigma_1^6/3
\end{bmatrix}\Bigg),$$

$$\begin{bmatrix}
    \Gamma_t' \\
    \Gamma_t
\end{bmatrix} \bigg |
\begin{bmatrix}
    \Gamma_{t-1}' \\
    \Gamma_{t-1}
\end{bmatrix}\sim \text{Normal}_2\Bigg(\begin{bmatrix}
    \Gamma'_{t-1} \\
    \Gamma_{t-1} + \sigma^2 \Gamma'_{t-1}
\end{bmatrix}, \begin{bmatrix}
    \sigma^2  & \sigma^4/2\\
   \sigma^4/2 &  \sigma^6/3
\end{bmatrix}\Bigg) \; t=2,...,T,$$

\noindent where $\sigma^2$ scales time. We use a log-normal prior on $\sigma$, where the parameters are chosen based on expected week-to-week (day-to-day) variation in $R_t$.

While the joint density of $\mathbf{\Gamma}$ can still be expressed as a simple product of conditional densities, we must keep in mind that we are now modelling an additional vector of parameters: $\mathbf{\Gamma'}$, the time derivative of $\mathbf{\Gamma}$. We then have the joint density of both $\mathbf{\Gamma}$ and $\mathbf{\Gamma'}$:

$$P(\mathbf{\Gamma}, \mathbf{\Gamma'}\mid \sigma) = P(\Gamma_1, \Gamma_1'\mid \mu_1', \mu_1, \sigma_1)\prod_{t=1}^T P(\Gamma_t, \Gamma_t' \mid \Gamma_{t-1}, \Gamma_{t-1}', \sigma).$$

\subsubsection{Hilbert Space Approximation to Gaussian Process with Matérn Kernel (HSGP)}

The Hilbert Space Gaussian Process approximation (HSGP), developed by \citet{Solin_Sarkka_2019}, can approximate a Gaussian process with a stationary kernel function. Their method is a part of a class of methods that aim to approximate the covariance matrix with a lower rank matrix via approximate eigendecomposition. The general idea is to treat the stationary covariance function as a pseudo-differential operator and approximate it using Hilbert space methods that are used in partial differential equations. This computation scales linearly with the number of data points, as opposed to cubically as in the case of using the GP with its exact covariance function.

We include the HSGP prior in our comparison because it underpins the default smoothing prior used by EpiNow2 \citep{epinow2}. In EpiNow2, a multivariate normal prior is placed 
on the vector of first-order differences of the latent log reproduction number, 
$\mathbf{\Delta\Gamma} = (\Gamma_2 - \Gamma_1, \dots, \Gamma_T - \Gamma_{T-1})$, with mean zero and a covariance structure motivated by a Matérn $3/2$ kernel. This covariance is obtained via the HSGP approximation, which provides a low-rank representation of the Matérn $3/2$ structure on this finite set of differences. We will compare our models to EpiNow2 directly, but we will also use the HSGP prior with a Matérn 3/2 kernel under our likelihood directly on the latent function $\Gamma_t$ for comparability with the other priors in this study. Evaluating this approximate GP prior on the grid of observations gives

$$\mathbf\Gamma \overset{\cdot}\sim \text{Normal}_T(\mathbf{0}, K),$$

\noindent where covariance entries $K_{ij} \approx k(\mid t_i - t_j\mid),$ with

$$k(\Delta t) = \alpha^2 \bigg(1 + \frac{\sqrt{3}\Delta t}{\ell}\bigg)\exp\bigg(-\frac{\sqrt{3}\Delta t}{\ell}\bigg),$$

\noindent where $\ell>0$ represents the length scale and $\alpha > 0$ represents the magnitude of the kernel. Unlike the priors with a Markov property, this specification does not admit a simple product of low-order conditional densities; the implied joint density is determined by
the full covariance structure,

$$P(\mathbf{\Gamma}\mid\alpha,\ell)
  \propto |K|^{-1/2}\exp\!\left(-\tfrac12\mathbf{\Gamma}^T
  K^{-1}\mathbf{\Gamma}\right),$$
  
\noindent though the HSGP approximation represents this covariance in low rank and avoids inverting the full, exact covariance matrix.

When using the GP with exact Matérn covariance, priors would need to be placed on the marginal variance $\alpha^2$ and the length scale $\ell$ which controls the smoothness, or the strength of correlation at distances $\Delta t$. Approximating this Matérn GP with the HSGP necessitates setting a few more hyperparameters: the number of basis functions to use in the approximation and the boundary condition. \citet{Riutort-Mayol_etal_2022} give practical guidance on choosing these settings, a two-phase process which involves iteratively guessing the length scale, setting the boundary condition and basis functions based on a provided set of equations, running the model and checking conditions, then iteratively increasing the number of basis functions, running the model, and checking on the stability of the predictive accuracy. Though this guidance is very helpful to the user, the inconvenience of the process along with the difficulty of the underlying mathematics may be enough to deter practitioners from using the HSGP. Practitioners may alternatively be inclined to apply the HSGP prior without properly choosing settings, leading to invalid inference. It is reasonable to believe that many users of EpiNow2 may rely on the default settings for the HSGP prior, rather than making modifications based on a thorough evaluation of the approximation accuracy.

\subsection{Posterior Inference}

For models using the RW1, OU, RW2, and HSGP priors, the posterior distribution of incidence, log effective reproduction number, and other nuisance parameters can be represented by

$$P(\mathbf{ I},\mathbf{ \Gamma}, \bm{\xi}, \rho, \kappa, \nu, \lambda \mid \mathbf{ O}) \propto \underbrace{P(\mathbf{ O}\mid \mathbf{ I}, \rho,\kappa)}_{\text{observed cases}}\underbrace{P(\mathbf{ I}\mid \mathbf{ \Gamma},\nu,\lambda)}_{\text{latent cases}} \underbrace{P(\mathbf{ \Gamma}\mid \bm{\xi})}_{\log R_t\text{ prior}} \underbrace{P(\bm{\xi},\rho,\kappa,\nu,\lambda)}_{\text{other priors}},$$

\noindent where $\bm{\xi}$ is the vector of parameters that the given $\mathbf{\Gamma}$ prior relies on. This posterior form needs to be altered for the models which use the IBM prior, since an additional latent process is modelled and the state space is therefore larger:

$$P(\mathbf{ I},\mathbf{ \Gamma}, \mathbf{\Gamma'}, \bm{\xi}, \rho, \kappa, \nu, \lambda \mid \mathbf{ O}) \propto \underbrace{P(\mathbf{ O}\mid \mathbf{ I}, \rho,\kappa)}_{\text{observed cases}}\underbrace{P(\mathbf{ I}\mid \mathbf{ \Gamma},\nu,\lambda)}_{\text{latent cases}} \underbrace{P(\mathbf{ \Gamma}, \mathbf{\Gamma'}\mid \bm{\xi})}_{\text{IBM prior}} \underbrace{P(\bm{\xi},\rho,\kappa,\nu,\lambda)}_{\text{other priors}}.$$

To approximate these posterior distributions, we use Hamiltonian Monte Carlo with the No-U-Turn Sampler (NUTS) as implemented in the R package \texttt{rstan} \citep{stan_dev_team, hoffman_gelman_2014}. All code needed to reproduce the results below is available at \url{https://github.com/jessalynnsebastian/gauss-markov-rt}.

\section{Results}

\subsection{Application to Simulated Respiratory Virus Circulation}

\subsubsection{Simulation Protocol}

We simulated circulation of a respiratory virus from a stochastic Susceptible-Exposed-Infected-Removed-Susceptible (SEIRS) model in R. The SEIRS model is a compartmental model, a class of models that divides a population into compartments between which individuals can transition according to a continuous time Markov chain with certain rate parameters. In this case, the population is divided into susceptible, exposed (infected but not yet infectious), infectious, and removed (recovered, but not yet susceptible again). SEIRS dynamics are appropriate here because they capture waning immunity, which is important for realistically simulating case trajectories consistent with respiratory virus circulation. The use of SEIRS, rather than a renewal equation–based simulation, is also relevant: while SEIRS generates realistic epidemic dynamics, these dynamics differ from the renewal equation models we fit, meaning that we introduce a realistic degree of model misspecification. This simulation design allows us to benchmark each prior against known ground-truth $R_t$ trajectories under realistic epidemic dynamics.

The simulated scenario lasted one year. The time-varying basic reproduction number, $R_0(t)$, which gives the average number of individuals an infected person will go on to infect assuming an entirely susceptible population, was given a fixed trajectory. True $R_t$ for each simulation is obtained by multiplying $R_0(t)$ with the time-varying proportion of susceptible individuals in that simulation. Observed case data was generated from a negative binomial distribution with mean proportional to the number of transitions from the E to the I compartment. Note that, as discussed in Section \ref{sec:likelihood}, this assumes no delay between the transition to the infectious stage and observation of the case; the only delay in observation comes from the time between being infected and becoming infectious. We simulated 100 realisations from this scenario to compare models across a variety of stochastic variations in individual scenarios. Simulation parameters are given in Table \ref{table:seirs_params}.

The five models described in Section \ref{sec:rt_priors} were fit to each of these 100 simulated outbreaks and their $R_t$ inference was compared. We additionally fit EpiNow2, using its default prior (HSGP on successive differences in $\Gamma_t$), as it is a commonly used method for $R_t$ inference. It should be noted that their likelihood is very similar to the likelihood used by our other models, though with a few differences. Details on EpiNow2 implementation can be found in Appendix \ref{sec:epinow2}.

We assessed the relative performance of the models using several metrics: the mean absolute deviation (MAD), envelope, and mean credible interval width (MCIW). Taking $\hat R_t$ to be the posterior median of $R_t$, with a time series of length $T$, these summary statistics are defined as

\begin{align*}
\text{MAD} &= \frac{1}{T} \sum_{i=1}^T \mid \hat R_i - R_i \mid, \\
\text{envelope} &= \frac{1}{T} \sum_{i=1}^T \mathbbm{1}\{R_i \in (\hat R_{i, p_1}, \hat R_{i, p_2})\}, \\
\text{MCIW} &= \frac{1}{T} \sum_{i=1}^T (\hat R_{i, p_2} - \hat R_{i, p_1}),
\end{align*}

\noindent where $\mathbbm{1}\{\cdot\}$ is the indicator function and $\hat R_{i, p_j}$ represents the $p_j$th quantile of the posterior distribution of $R_i$. This gives us a measure of distance between the point estimate and the truth (MAD), coverage of BCIs (envelope), and precision of inference (MCIW). Here, we take quantiles that yield envelope and MCIW metrics for 80\% ($p_1 = 0.1, \; p_2 = 0.9$) and 95\% ($p_1=0.025, \; p_2 = 0.975$) BCIs.

\subsubsection{Simulation Results}

All models were fit using \texttt{rstan} v.2.32.7 \citep{stan_dev_team}. For each model, we ran 4 chains of 6000 iterations each and dispensed the first 2000 as burn-in. Each model was fit once to each of the 100 simulated datasets.

Table \ref{table:freq_metrics} summarises the retrospective performance of the six models across 100 simulated outbreaks. The RW1 prior behaves as expected, producing comparatively wide posterior intervals that almost always contain the truth. Its 95\% envelopes approach 1 (99.4\% on average), and this high coverage comes at the cost of precision: its average MCIW is 1.5--2 times as large as that of many of the other models. RW1's 80\% envelopes cover at a rate even further above nominal, at about 93.8\%, and its 80\% BCIs are correspondingly wide, at over double the width of those for some of the other models. The RW1 prior allows the posterior to cover the true $R_t$ trajectory but dilutes the usefulness of the inference. The performance of the OU model is quite similar, with high 95\% envelopes (99.3\% on average) and similar MAD and MCIW to RW1. Though the epidemiological significance of the OU prior's mean reversion made it seem appealing, it appears to suffer from the same problems as the RW1, and with unneccessary extra complexity.

\begin{table}[htbp]
  \centering
  \begin{tabular}{lrrrrrr}
  \toprule
    & \multicolumn{2}{c}{Envelope} & \centering MAD & \multicolumn{2}{c}{MCIW} \\
    \cmidrule(lr){2-3} \cmidrule(lr){5-6}
    & 95\% BCI & 80\% BCI & & 95\% BCI & 80\% BCI \\
  \midrule
  RW1 & 99.4\% (1.3\%) & 93.8\% (4.6\%) & 0.060 (0.011) & 0.445 (0.023) & 0.286 (0.015) \\
  OU & 99.3\% (1.3\%) & 93.7\% (4.8\%) & 0.062 (0.011) & 0.463 (0.025) & 0.297 (0.016) \\
  RW2 & \textbf{95.3\% (4.2\%)} & 82.5\% (8.0\%) & 0.047 (0.009) & 0.247 (0.014) & 0.158 (0.009) \\
  IBM & 93.8\% (5.1\%) & \textbf{79.2\% (9.6\%)} & 0.045 (0.009) & 0.221 (0.013) & \textbf{0.141 (0.008)} \\
  HSGP & 92.9\% (7.5\%) & \textbf{79.2\% (12.8\%)} & 0.056 (0.017) & 0.272 (0.016) & 0.175 (0.010) \\
  EpiNow2 & 97.1\% (5.1\%) & 86.5\% (8.1\%) & \textbf{0.038 (0.009)} & \textbf{0.220 (0.014)} & \textbf{0.141 (0.009)} \\
  \bottomrule
  \end{tabular}
  \caption{Mean (SD) of three summaries for each model: the envelope (proportion of time the true curve lies inside the 95\% or 80\% intervals), the median absolute deviation between point estimates and truth (MAD), and the mean width of the 95\% and 80\% credible intervals (MCIW).}
  \label{table:freq_metrics}
\end{table}

By contrast, the IBM and RW2 priors achieve a better balance between coverage and precision. Both yield coverage close to nominal (93.8\% and 95.3\%, respectively, for 95\% intervals; 79.2\% and 82.5\% for 80\% intervals) while producing substantially narrower BCIs (95\% MCIW 0.22–0.25, 80\% MCIW 0.14-0.16) and lower MADs than RW1 or OU. This suggests that their added smoothness relative to RW1 is effective in constraining the posterior without sacrificing coverage.

The two HSGP-based approaches also perform well. The HSGP prior produces intervals narrower than RW1 and OU, with average coverage near nominal (92.9\% for 95\% BCIs, 79.2\% for 80\% BCIs). EpiNow2 performs especially well in this setting: it achieves the highest average coverage among the smoother priors (97.1\% for 95\% BCIs, 86.5\% for 80\% BCIs) and the lowest MAD overall (0.038), with interval widths comparable to IBM.

\begin{figure}[htbp]
    \centering
    \includegraphics[width = 0.9\textwidth]{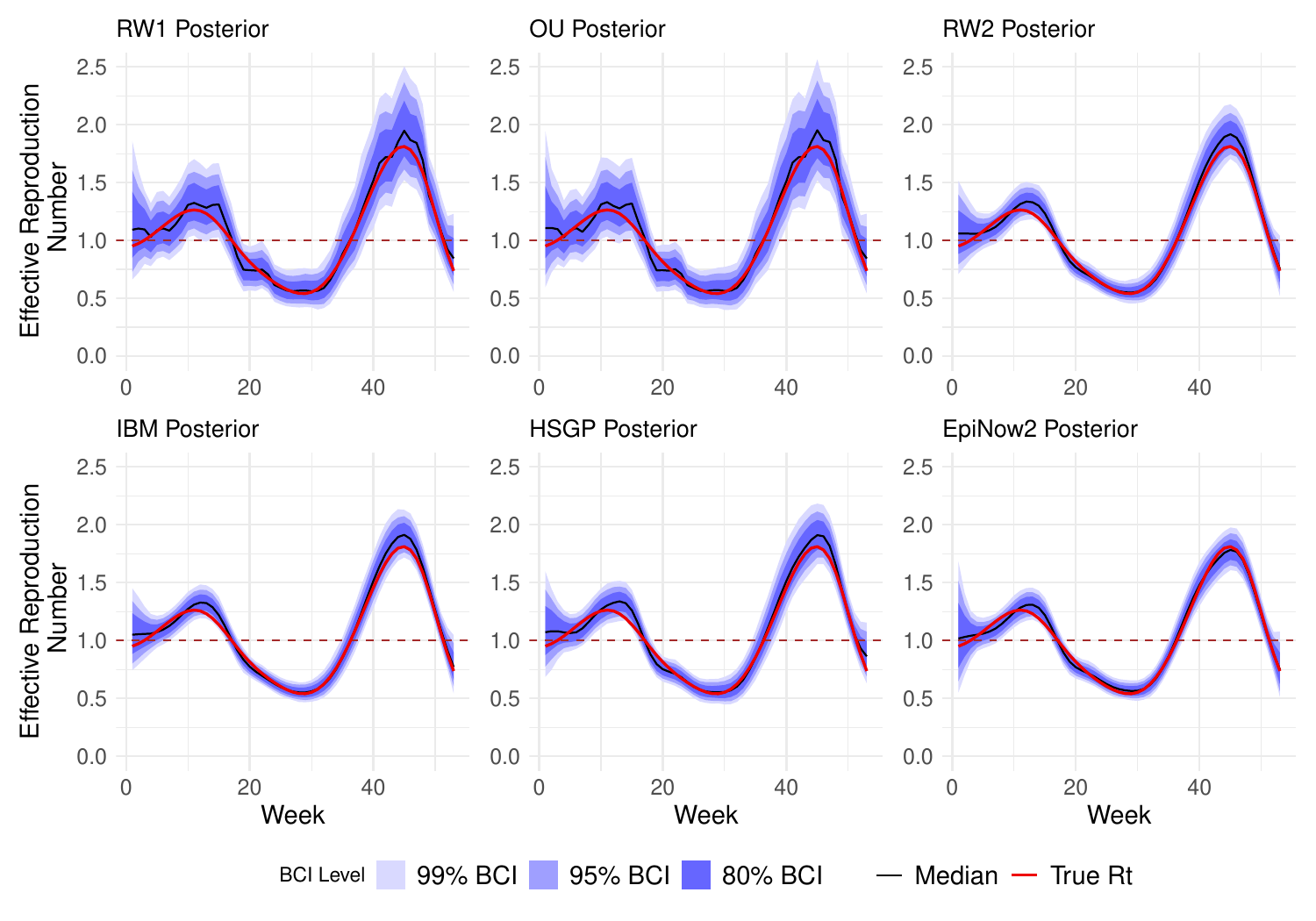}
    \caption{Example posterior $R_t$ plots by prior for one simulated outbreak. True simulated $R_t$ is given in red, and 99\%, 95\%, and 80\% posterior Bayesian credible intervals are shaded in blue. The brown dashed line emphasises the critical value 1, where $R_t>1$ implies the outbreak is growing, and $R_t<1$ implies the outbreak is shrinking.}
    \label{fig:sim_posteriors}
\end{figure}

Figure \ref{fig:sim_posteriors} illustrates these differences for an example simulated outbreak. The RW1 and OU models yield broad, less informative intervals, whereas the IBM and RW2 priors constrain uncertainty more effectively while still following the trajectory of the truth. The HSGP and EpiNow2 models similarly provide smooth posterior trajectories with reasonably narrow intervals.

Table \ref{table:cpu_time}, which gives CPU time comparisons for all models, is  reported in the Appendix. Computation time was modest overall, with average runtimes between 3 and 6 minutes per fit. RW1, OU, and HSGP were the fastest, followed by RW2, with IBM requiring slightly more time on average. These results demonstrate that the Markov IBM and RW2 combine the ease and speed of RW1 with inferential performance comparable to more complex Gaussian process approximations like HSGP, while remaining straightforward to implement.

When inferring $R_t$ in real time, modellers are generally most interested in the most recent $R_t$ inference as opposed to retrospective inference, since current values are the most relevant for informing policy decisions. To reflect this, we evaluated the real-time performance of the six priors separately from their retrospective performance.

In reality, real-time inference of $R_t$ is hindered by right truncation of case data; i.e., cases that have occurred by time $t$, but have not yet been reported and recorded. Estimating and accounting for delays is an active area of research with many creative approaches \citep{seaman_estimating_2022}. For the purposes of this project, since we are primarily interested in the relative performances of only the priors of $R_t$, we treat right truncation as a separate problem outside the scope of this paper and assume no reporting delays are present. Our goal here is to isolate the behaviour of the prior itself, not the method used for delay correction, and omitting reporting delay handling allows us to avoid conflating two distinct sources of uncertainty.

To compare the real-time performance of the six models, we first apply each model to the first ten weeks of one of the simulated datasets. Then, we iteratively extend the dataset by adding one additional week of data, applying all models at each step to the progressively longer time series. At each iteration, only the posterior samples of $R_t$ for the most recent time point are retained.

In addition to computing the metrics we used in the retrospective analyses, we define a decision ``score'' motivated by a hypothetical policy setting in which real-time decisions are based on posterior credible intervals for $R_t$. The decision score, as we define it, is the proportion of weeks in which the credible interval correctly indicates whether transmission is increasing or decreasing,

$$\text{Decision Score} = \frac{1}{T_N}\sum_{i=0}^{N} \left ( \mathbbm{1}\{\hat R_{T_i, p_1} > 1\}\mathbbm{1}\{R_{T_i} > 1\} + \mathbbm{1}\{\hat R_{T_i, p_2} < 1\}\mathbbm{1}\{R_{T_i} < 1\}\right ),$$

\noindent where $N$ is the number of weeks for which real-time inference is considered, and $T_i$ is the last week of the $i$th observational period. This measure does not capture the full distributional uncertainty, but instead provides a concise summary of the model’s performance for threshold-based decision making in real time.

\begin{figure}[htbp]
    \centering
    \includegraphics[width = 0.9\textwidth]{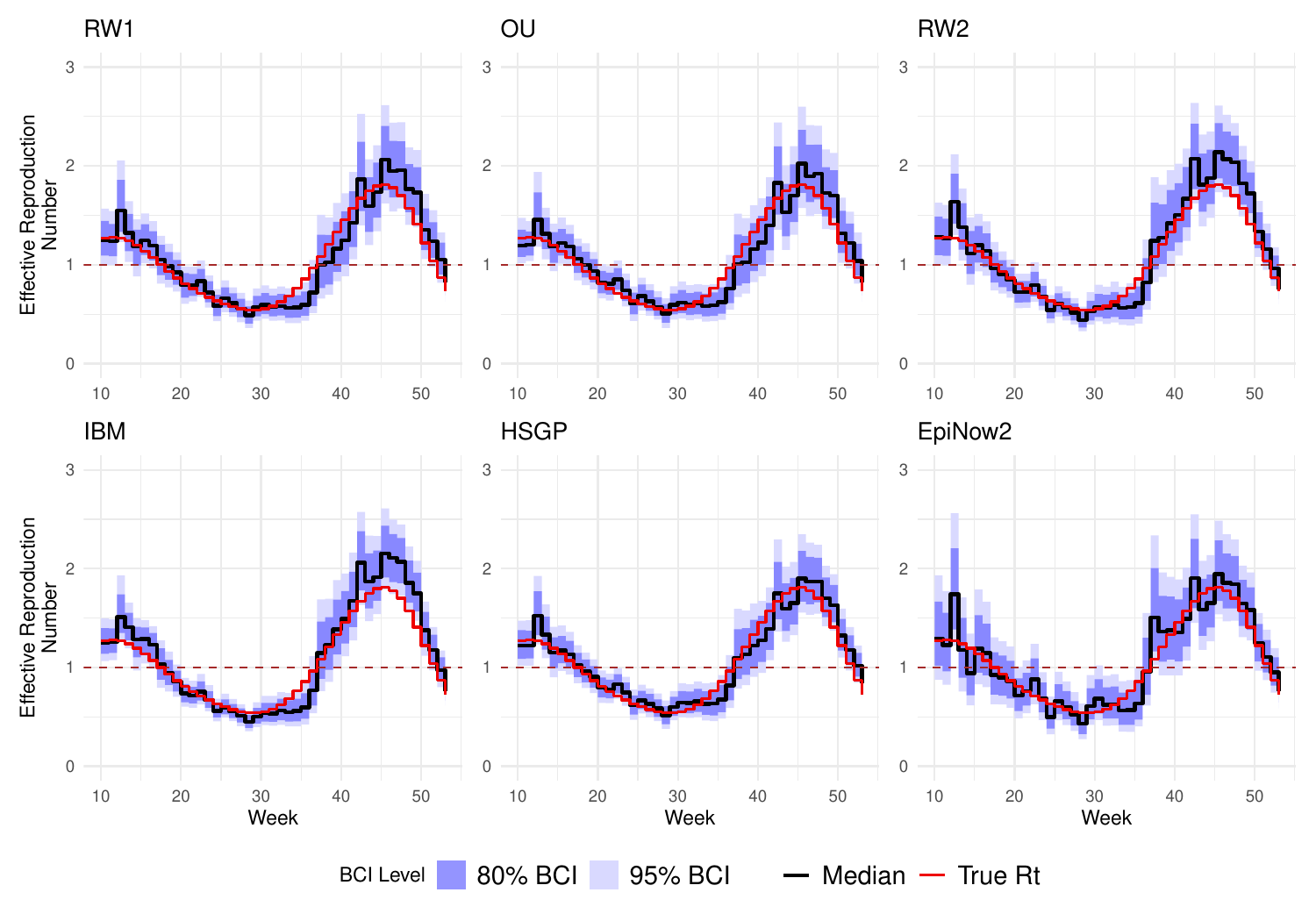}
    \caption{Visual representation of real-time inference by the six models. True $R_t$, posterior medians, and corresponding credible intervals are represented with step functions to emphasise that each estimate and its uncertainty correspond to a different iteration of the model, each relying on different data. The truth is given in red.}
    \label{fig:realtime_sims}
\end{figure}

Figure \ref{fig:realtime_sims} and Table \ref{table:realtime_sims} summarise the models’ real-time performance when applied iteratively to progressively longer simulated datasets as described above. All six priors achieved coverage close to or above nominal, with several models (HSGP, OU, and EpiNow2) reaching 100\%. The relevant differences are in precision and, by extension, decision-making.

The HSGP prior stood out with the lowest mean absolute deviation (MAD = 0.091), high coverage (100\% for 95\% BCIs, 84.1\% for 80\% BCIs), and the narrowest credible intervals (0.49 for 95\% BCIs, 0.31 for 80\% BCIs), reflecting its ability to deliver both accurate point estimates and efficient uncertainty quantification in real time. IBM performed comparably, though not quite as outstandingly, with coverage near nominal (93.2\% for 95\% BCIs, 75.0\% for 80\% BCIs), credible intervals of the same size as the HSGP's (0.50 for 95\% BCIs, 0.31 for 80\% BCIs), and the highest decision score of all models (84.1\%).

Retrospectively, posterior credible intervals produced by models using the RW1 and OU priors were the widest overall, leading to over-coverage and reduced precision. In the real-time setting, however, their intervals were still relatively wide but not nearly as extreme, and this translated into very strong coverage (98–100\% for 95\% BCIs, 77-82\% for 80\% BCIs) and reasonably high decision scores. In other words, the drawbacks of RW1 and OU are much more prominent when we are interested in inference on the entire $R_t$ trajectory, whereas in real-time monitoring they perform somewhat competitively.

EpiNow2 also presented an interesting contrast here; it achieved perfect coverage across simulations but with by far the widest intervals across models (0.72 for 95\% BCIs, 0.46 for 80\% BCIs). This suggests that the EpiNow2 approach of differencing, where the HSGP prior is placed on successive differences in $\Gamma_t$, is better-suited for retrospective inference than for real-time inference. The real-time performance of RW2 in some sense seconds this suggestion (as it is also a method of differencing: the RW2 is just the RW1 on differences in $\Gamma_t$), with real-time credible intervals wider than the OU or RW1 despite much higher precision in the retrospective study.

\begin{table}[htbp]
  \centering
  \begin{tabular}{lrrrrrrrr}
  \toprule
    & \multicolumn{2}{c}{Real-Time Coverage} & \centering MAD & \multicolumn{2}{c}{MCIW} & \multicolumn{2}{c}{Decision Score}\\
    \cmidrule(lr){2-3} \cmidrule(lr){5-6} \cmidrule(lr){7-8}
    & 95\% BCI & 80\% BCI & & 95\% BCI & 80\% BCI & 95\% BCI & 80\% BCI \\
  \midrule
  RW1 & 97.7\% & 77.3\% & 0.118 & 0.565 & 0.357 & 77.3\% & 88.6\% \\
  OU & 100.0\% & \textbf{81.8}\% & 0.114 & 0.553 & 0.347 & 81.8\% & 88.6\% \\
  RW2 & \textbf{95.5\%} & 77.3\% & 0.112 & 0.577 & 0.367 & 81.8\% & 90.9\% \\
  IBM & 93.2\% & 75.0\% & 0.119 & 0.498 & \textbf{0.313} & \textbf{84.1\%} & \textbf{93.1\%} \\
  HSGP & 100.0\% & 84.1\% & \textbf{0.091} & \textbf{0.491} & \textbf{0.313} & 79.5\% & 90.9\% \\
  EpiNow2 & 100.0\% & 86.4\% & 0.106 & 0.722 & 0.455 & 72.7\% & 84.1\% \\
  \bottomrule
  \end{tabular}
  \caption{Summary of real-time model performance for one simulated dataset. Refer to Table \ref{table:freq_metrics} for definitions of abbreviations. At each time point we evaluate only the most recent posterior $R_t$ inference, so there is only one value of each metric per time point. Because these quantities are not sampled across replicated datasets, there is no Monte Carlo variability in the sense used in the 100-dataset simulation study; therefore we do not report standard deviations in this table.}
  \label{table:realtime_sims}
\end{table}

Across both retrospective and real-time analyses, we see a consistent pattern. The RW1 and OU priors provide high coverage but at the cost of reduced precision. The HSGP delivers high coverage and precision retrospectively and in real time, but is more challenging to implement. IBM combines the strengths of both approaches: it is computationally efficient, straightforward to implement, and produces posterior inference that is precise, well-calibrated, and effective for monitoring transmission dynamics both retrospectively and in real time.

\subsection{Application to Data from the Counties of Los Angeles and San Francisco, California}
\label{sec:realdata}

We compare the results of applying the six models described in Section \ref{sec:rt_priors} to data from two California counties: Los Angeles and San Francisco. These two counties are major population hubs in the state, but their populations and densities are quite different. Los Angeles is the most populated county in the country, at around 10 million residents, who are spread over a wide area (over 12,000 square kilometers). San Francisco has a much smaller population, but it is much denser: all 800,000 residents live within about 120 square kilometers. The data used in these analyses were time series of cases (ordered by the date of their polymerase chain reaction, or PCR, test) reported between mid-June of 2020 and mid-February of 2021, and available at the California Open Data portal \citep{ca_open_data_portal}.

To apply our renewal equation-based models to real case data, we need to choose a generation time distribution, a latent period distribution to use as a delay distribution, and priors for case overdispersion $\kappa$, ascertainment rate $\rho$, and the mean-variance relationship for latent incidence $\nu$.

The choice of generation time distribution is a challenge, partly because it is difficult to estimate, partly because different variants of the same pathogen may have different intrinsic generation times, and partly because in reality, generation time would change with depletion of susceptible individuals in the population. In renewal equation-based models, it is known that longer assumed mean generation times lead to greater amplitude of estimates away from 1; that is, $R_t$ estimates larger than 1 are increased and $R_t$ estimates less than 1 are decreased \citep{wallinga_generation_2007}. Based on the work of \citet{Hart_2022}, \citet{Park_2023}, and \cite{Manica_2022}, we use a Gamma generation time distribution with a mean of 4.6 days and a standard deviation of 1.2 days. For the latent period distribution, we take the estimate of \citet{xin_latent_2021}, which has a mean of 5.5 days and a standard deviation of 2.5 days. Changing the mean latent period primarily affects the timing of the $R_t$ inference, i.e., translates it forward or backward in time. We do not anticipate significant effects of the choice of these distributions of the relative differences between the $R_t$ posteriors for the 6 models.

The choice of $\kappa$ is briefly discussed in Section \ref{sec:likelihood}, and is further discussed in Appendix \ref{sec:kappa}. A complete list of priors can be found in Table \ref{table:priors}.

\subsubsection{Retrospective Analyses}

\begin{figure}[htbp]
    \centering
    \includegraphics[width = \textwidth]{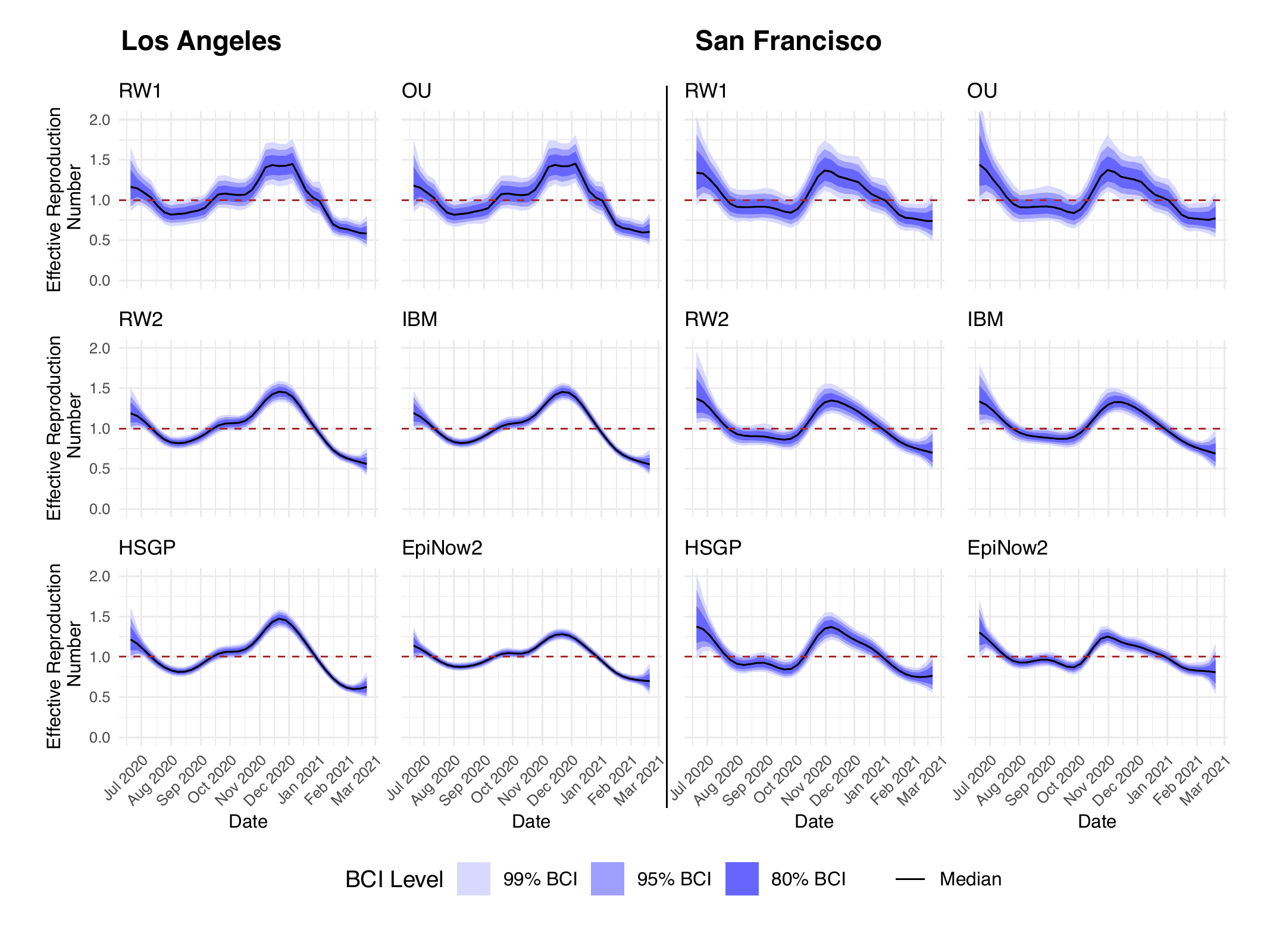}
    \caption{Posterior $R_t$ inference for Los Angeles County and San Francisco County from June 2020 to February 2021 under six priors. Solid black lines show posterior medians, dashed brown lines denote the critical value 1, and shaded blue bands represent 80\%, 95\%, and 99\% Bayesian credible intervals.}
    \label{fig:realdata_retro}
\end{figure}

Figure \ref{fig:realdata_retro} summarizes the retrospective analyses using all six models for each of Los Angeles and San Francisco counties. All models captured the broad features of the COVID-19 pandemic in Los Angeles during 2020--2021, including a period of relatively stable transmission through summer and early fall, followed by a steep increase in $R_t$ during November and December 2020 that peaked at around 1.5 before declining in early 2021. Consistent with the simulation study, RW1 and OU produced the widest 95\% credible intervals, often spanning values well above and below one, while IBM and RW2 yielded narrower intervals that tracked a similar trajectory. The HSGP and EpiNow2 estimates were similar in smoothness to IBM and RW2 and produced similarly narrow credible intervals, to varying degrees. 

In San Francisco, where case counts were much smaller, the models again agreed on the major features; relatively modest transmission through summer and fall is seen, followed by a rise above one in December 2020 and a decline in early 2021. Differences across priors were more pronounced than in Los Angeles. RW1 and OU yielded very wide intervals that at times obscured the shape of the trajectory. IBM, RW2, HSGP, and EpiNow2 are more certain, to varying degrees.

All models recovered the winter 2020--2021 surge, but EpiNow2 produced a visibly smaller peak than the other models. This could be related to a number of factors, including higher-order smoothing and a deterministic renewal specification (see Appendix \ref{sec:epinow2}). A similar, though less extreme, difference is noticeable in the simulation study results. Importantly, despite the lower peak height for EpiNow2, all models agreed that $R_t$ exceeded 1 during the surge. Ignoring differences in uncertainty, the priors gave a consistent narrative of transmission dynamics in this large-population setting.

\subsubsection{Real-Time Analyses}

\begin{figure}[htbp]
    \centering
    \includegraphics[width = \textwidth]{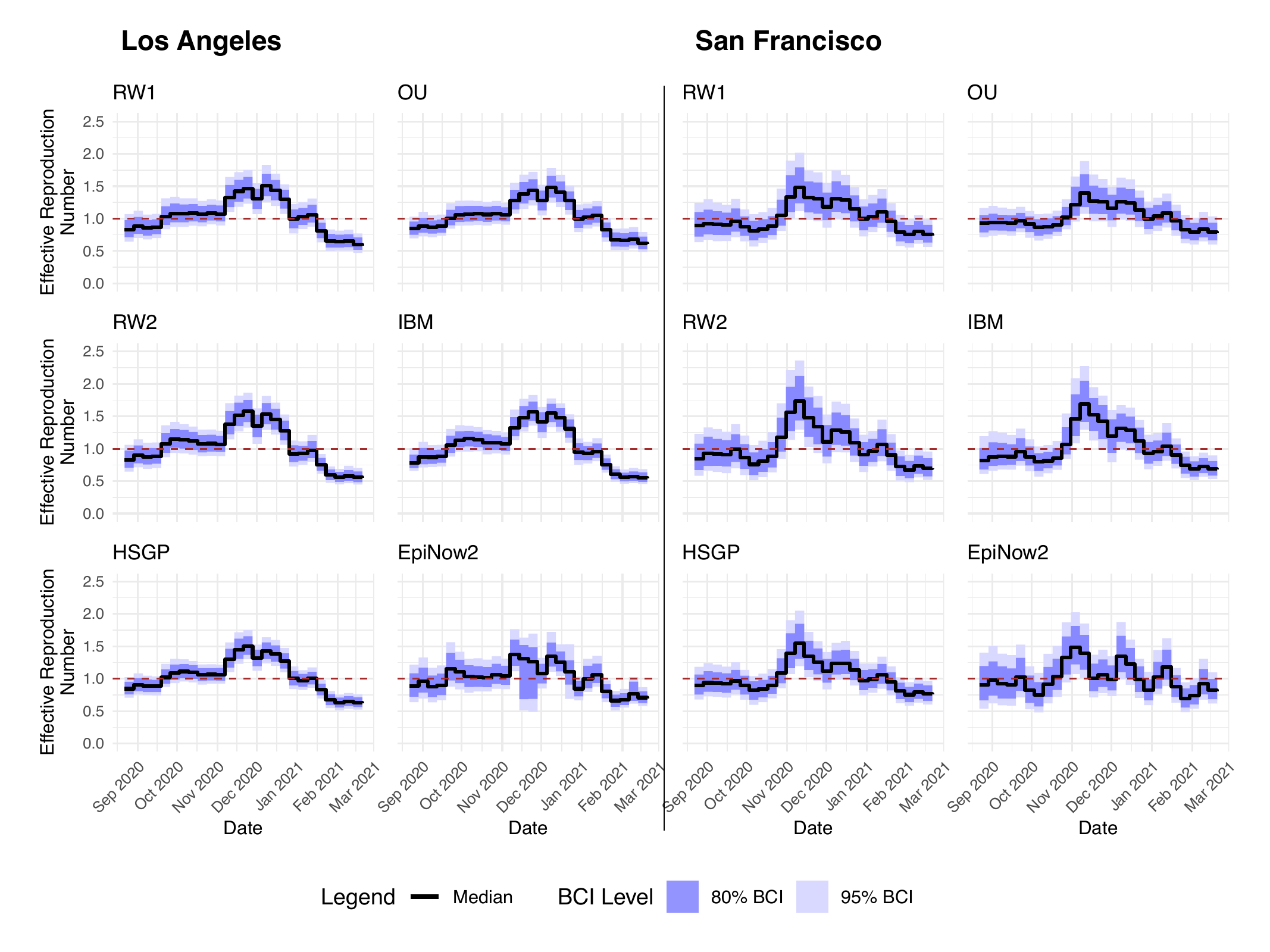}
    \caption{Real-time inference of $R_t$ for Los Angeles County and San Francisco County, updated weekly through February 2021. Solid black lines show posterior medians, dashed brown lines denote the critical value 1, and shaded blue bands represent 80\% and 95\% Bayesian credible intervals.}
    \label{fig:realdata_realtime}
\end{figure}

Figure \ref{fig:realdata_realtime} gives a visual summary of the real-time inference of the six models in Los Angeles County and San Francisco County. In Los Angeles, the real-time analyses showed general consistency across all models, with the exception of EpiNow2, which maintains the more uncertain real-time behaviour observed in the simulation study. Each other model’s 95\% credible interval for $R_t$ rose above one by late November 2020 and remained there throughout December, clearly signaling growing transmission. The differences that were more visible in retrospective fits were much less pronounced in the real-time setting. In practice, a policymaker relying on any of these priors would have received the same actionable signal, with the exception of possibly the EpiNow2 model: transmission was increasing rapidly and $R_t$ was well above threshold. This suggests that in large populations with high case counts, the choice of prior has less influence on real-time inference.

The lower case counts in San Francisco made the decision problem across priors more interesting. IBM, RW2, and HSGP produced credible intervals that crossed the 1.0 threshold quickly in December, supporting early inference that the outbreak was growing. RW1 and OU again yielded wider intervals, but during the surge their posteriors still lay mostly above one, which would have led to the same decision albeit with less certainty. EpiNow2 models, by contrast, produced trajectories with wider credible intervals that often overlapped the critical value 1 for longer time periods.

\section{Discussion}

We compared several Gaussian process priors with Markov properties for modelling $R_t$ and evaluated their performance relative to widely used alternatives, including current state-of-the-art. In retrospective analyses, the IBM prior, HSGP approximation to a Gaussian process prior with a Matérn kernel, and RW2 prior for $R_t$ produced well-calibrated posterior inference that balanced coverage of the true simulated $R_t$ trajectory with precision. The OU process prior and the widely-used RW1 prior over-covered the truth with unnecessarily wide posterior credible intervals. Popular epidemic modeling package EpiNow2 performed well, with conservative inference that still managed to outperform other models in terms of precision. Interestingly, these results did not hold true when the focus is shifted to real-time $R_t$ inference, where only the most recent $R_t$ inference is of interest. In real time, while IBM and HSGP maintained near-nominal coverage and high precision, EpiNow2 and RW2 models suffered slightly: though their coverage was nominal (or even above nominal, in the case of EpiNow2), their real-time inference had the lowest and second-lowest precision, respectively, across the models in this study. Both RW1 and OU performed comparably to, if not slightly better than, the RW2 in this setting, and far outperformed EpiNow2 in terms of precision. This is a complete reversal from the retrospective setting. The precise real-time inference using IBM and HSGP would have provided more information about transmission dynamics in the midst of an outbreak than the inference of EpiNow2, RW2, OU, and RW1.

A key contribution of this work is to demonstrate the practical value of representing IBM jointly with its corresponding Brownian motion, both as a prior for $R_t$ specifically and for Bayesian smoothing in general. The IBM prior yields performance on par with HSGP, but using the Markov formulation it is conceptually simpler and only modestly slower to fit; compared with exact Matérn Gaussian processes, it is substantially more efficient. The Markov representation of IBM also lends itself easily to extensions that are either not possible or considerably more complex under other priors. For example, to allow IBM to be locally adaptive (adaptive to changes in the smoothness of the underlying function), we could let the parameter $\sigma$ vary with time, similar to \citep{yue_bayesian_2014}. RW2, IBM's discrete-time counterpart, can be made locally adaptive, but because it is defined only on a regular grid, extending it to allow a smoothly varying notion of local roughness is more cumbersome \citep{lindgren_second-order_2008, faulkner_locally_2018}. The continuous-time formulation of IBM makes this type of extension more natural. A number of locally adaptive smoothing methods have been developed for time series analysis \citep{faulkner_locally_2018, schafer_locally_2023} with applications in economics, epidemiology, astronomy, and obstetrics \citep{elagin_locally_2008, faulkner_horseshoebased_2020, politsch_trend_2020, elovitz_cervicovaginal_2019}. A locally adaptive IBM prior could provide a conceptually straightforward and computationally tractable addition to this toolkit.

One limitation of this paper is that we only considered renewal models. Priors may behave differently under alternative model structures, such as ODE-based compartmental models \citep{kermack_contribution_1927, keeling_modeling_2008, tang_review_2020} or branching process formulations \citep{ball_strong_1995, blumberg_inference_2013}, for example. This is also the case for other types of data streams: our simulations in this study only address reported case data, but $R_t$ can also be inferred from hospitalization data and/or wastewater pathogen concentration data, among others \citep{sherratt_exploring_2021, huisman_wastewater-based_2022, goldstein_wastewater_2024}. It is not entirely clear whether our results in this paper will extend directly to other types of epidemic models or data streams. The models used in this paper also inherit the usual limitations of discrete-time renewal approaches. Renewal models assume a fixed and known generation time, but in reality generation times vary and can and do change over the course of an epidemic \citep{xu_assessing_2023}. In addition, when significant probability mass of the generation time distribution is less than the time interval between observations, renewal models necessarily treat adjacent time points as close to independent, making it increasingly difficult to identify $R_t$ as the generation time gets shorter relative to the time step. Nonetheless, it is encouraging that in our simulation study, the renewal models performed well overall, a finding reflected in many other studies \citep{gostic_practical_2020, nash_real-time_2022, bhatt_semi-mechanistic_2023}.

Our findings emphasise that the choice of prior on $R_t$ meaningfully shapes posterior inference and should not be treated only as a technical afterthought. Priors are a tool for embedding epidemiological knowledge into models, and careful prior selection is an essential component of robust $R_t$ inference. The widely used RW1 prior remains a reasonable baseline when data are highly informative, but our results show that IBM yields more precise inference in noisier settings with only modest increase in computational cost. IBM performs comparably to Gaussian process approximation methods such as HSGP, but it is easier to implement and extend, and it is substantially faster than an exact Matérn formulation. These features make IBM a promising choice for practical $R_t$ inference.

\section*{Acknowledgments}
This work was supported in part by the National Institutes of Health grant R01-AI170204.






\pagebreak

\appendix
\section*{Appendices}
\section{EpiNow2 Comparison Details}
\setcounter{table}{0}
\renewcommand{\thetable}{A\arabic{table}}
\setcounter{figure}{0}
\renewcommand{\thefigure}{A\arabic{figure}}
\label{sec:epinow2}

EpiNow2 estimates infections and the effective reproduction number $R_t$ via a discrete renewal model, similar to the models defined in Section \ref{sec:likelihood}. Letting $I_t$ denote true incidence on day $t$ and $g_s$ the discretised generation time probability mass for lag $s$:

$$I_t = R_t\sum_{j=-n}^{t-1} g_{t-j} I_j,$$

\noindent where the most relevant difference as compared to the incidence model used in this paper is the use of deterministic incidence, as opposed to stochastic. Here, $I_t$ is equal to $R_t\sum_{j=-n}^{t-1} g_{t-j} I_j$, whereas in the model described in \ref{sec:likelihood}, $R_t\sum_{j=-n}^{t-1} g_{t-j} I_j$ is the mean of a gamma distribution. This likely did not make much of a difference for the examples in this paper, but would make for noticeably different inference in small-population settings.

As in the other models described in this paper, observed counts $O_t$ are linked to infections by discrete convolutions with delay distributions and a likelihood (by default, negative binomial with optional day-of-week effects). The model includes an initial seeding window preceding the first observation in order to supply infection history for the renewal equation.

By default, EpiNow2 places a multivariate normal prior on the first-order 
differences of the latent log reproduction number,
$$\boldsymbol{\Delta\Gamma} 
= (\Gamma_{2} - \Gamma_{1},\; \Gamma_{3} - \Gamma_{2},\; \dots,\; \Gamma_{T} - \Gamma_{T-1}),$$

\noindent so that the vector of differences $\Delta \boldsymbol{\Gamma}$ has mean zero and a covariance structure motivated by a Matérn $3/2$ kernel. This covariance matrix is not used in full; instead, EpiNow2 uses a Hilbert space Gaussian process (HSGP) approximation to represent the Matérn covariance in low rank on a bounded domain. The initial reproduction number has a separate prior (by default, log-normal on the original scale, as with the models in Section~\ref{sec:likelihood}).

Probability densities for generation times and delays are discretised and passed as vectors. Because EpiNow2's discretised renewal equation excludes same-day transmission, the generation-time mass at day $0$ must be set to zero and the distribution renormalised over $s\geq 1$. We modelled $R_t$ at a \textit{weekly} time scale with $R_t$ piecewise-constant by week. While it is possible to treat EpiNow2's discrete time steps as weeks rather than days, the lack of support for probability mass at $g_0$ causes problems here for almost all respiratory viruses, since their generation times tend to be shorter than one week. Instead, to run EpiNow2 with weekly reports, we used their \texttt{fill\_missing()} function, which provides support for regular non-daily reporting patterns. This does, however, hinder EpiNow2 in terms of CPU time, since its latent time series to model are seven times as long as in the other models used in this paper.

To compare EpiNow2's daily inference to our weekly estimands, we aggregate their daily $R_t$ to weekly. Let $t$ index days and let $\mathcal{T}_W$ be the set of days in week $W$. Then $I_W = \sum_{t\in\mathcal{T}_W} I_t$.

We then define the weekly effective reproduction number $R_W$ to be the value that preserves the renewal equation when collapsing multiple days to a single weekly bin: $I_W \equiv R_W\sum_{t\in\mathcal{T}_W}\sum_{j=-n}^{t-1} g_{t-j} I_j$. Then $R_W$ is given by

$$R_W = \frac{I_W}{\sum_{t\in\mathcal{T}_W}\sum_{j=-n}^{t-1} g_{t-j} I_j} = \frac{\sum_{t\in T_W} I_t}{\sum_{t\in\mathcal{T}_W}\sum_{j=-n}^{t-1} g_{t-j} I_j} = \frac{\sum_{t\in T_W} R_t\sum_{j=-n}^{t-1} g_{t-j} I_j}{\sum_{t\in\mathcal{T}_W}\sum_{j=-n}^{t-1} g_{t-j} I_j}.$$

We compute $R_W$ per posterior draw using EpiNow2’s daily draws of $(R_t, I_t)$ and the same $g_s$. Under purely weekly data, EpiNow2’s daily $R_t$ within a week is weakly identified and driven primarily by the stochastic process prior; aggregating via the weekly $R_W$ yields a fair basis for comparison that respects the renewal equation on a weekly timescale.

For further details on EpiNow2's modelling framework and default settings, see \url{https://epiforecasts.io/EpiNow2/index.html}.

\FloatBarrier

\section{Simulation Study Parameters, Priors and Diagnostics}
\setcounter{table}{0}
\renewcommand{\thetable}{B\arabic{table}}
\setcounter{figure}{0}
\renewcommand{\thefigure}{B\arabic{figure}}
\subsection{Simulation Parameters}
\label{sec:params}

Weekly simulated case data was generated from transitions from the E to the I compartment of a stochastic SEIRS model using a negative binomial distribution. The parameters of the SEIRS model were:

\begin{table}[ht]
\centering
\begin{tabular}{c c c}
\toprule
Parameter & Interpretation & Value \\
\midrule
N & Population Size & 600,000 \\

$I_0$ & Initial Infectious & 50 \\

$\beta_t$ & \makecell{Time-Varying \\ Transmission Rate} & See figure below \\

$1/\sigma$ & \makecell{Mean latent \\ period (weeks)} & $4/7$\\

$1/\gamma$ & \makecell{Mean infectious \\ period (weeks)} & $7.5/7$ \\

$1/\omega$ & \makecell{Mean duration of \\ immunity (weeks)} & $12$ \\
\bottomrule
\end{tabular}
\caption{Parameters used in the SEIRS simulation.}
\label{table:seirs_params}
\end{table}

\begin{figure}[ht]
    \centering
    \includegraphics[width = 0.7\textwidth]{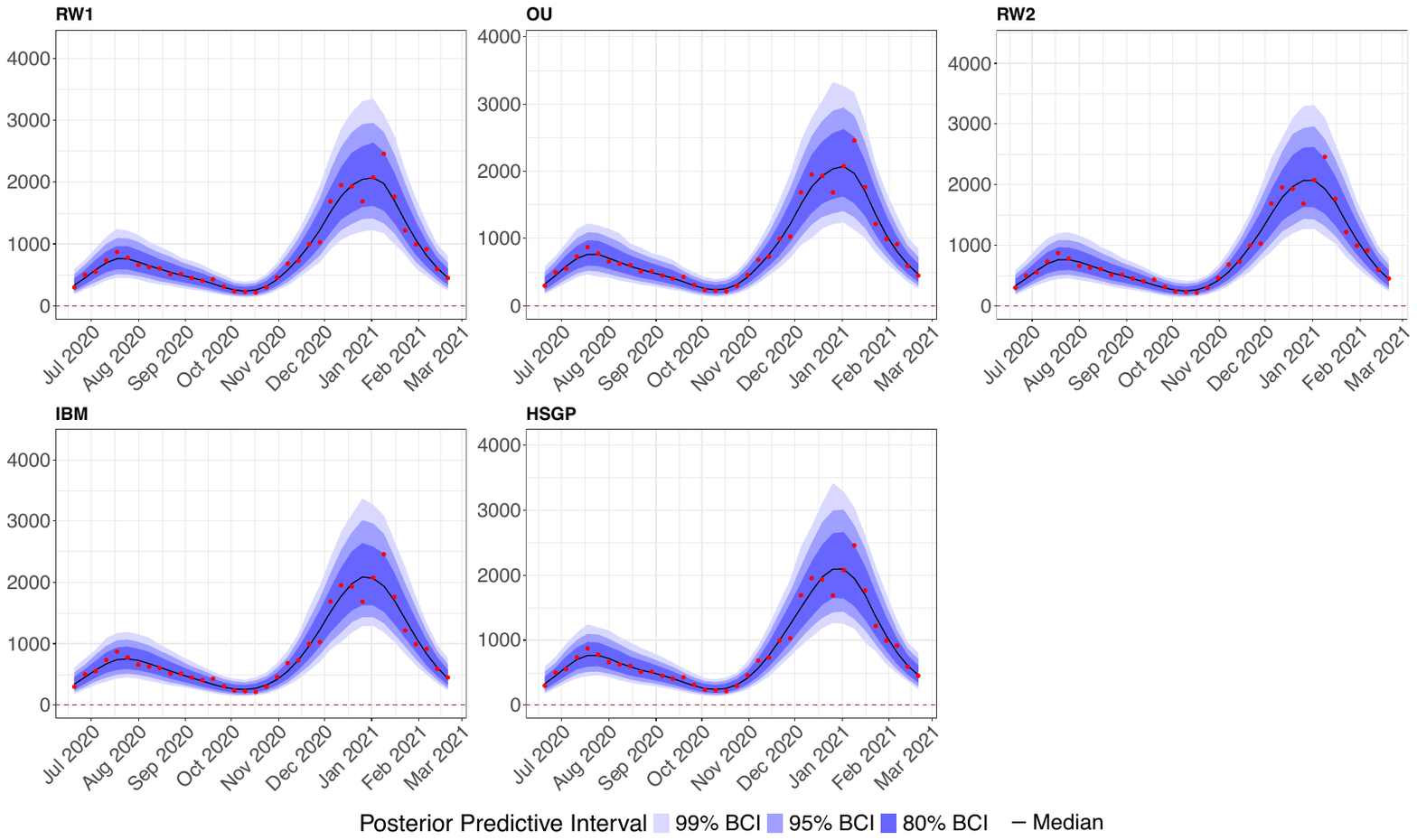}
    \caption{Time-varying transmission rate in the simulation study.}
    \label{fig:beta t}
\end{figure}

Transmission rate $\beta_t$ was given the time varying functional form displayed in Figure \ref{fig:beta t}.

Case data were generated from weekly counts of transitions from the E to the I compartment (E2I). Cases are generated by $O_t \sim \text{Negative Binomial} (\mu, \kappa)$, where the mean $\mu$ is the ascertainment rate $\rho$ multiplied by the E2I for time $t$, and $\kappa = 5$ is the overdispersion, where $\kappa\to\infty$ implies a Poisson distribution.

\subsection{Priors for Models Fit to Simulated Data}
\label{sec:priors_tbl}

\begin{table}[t]
\centering
\begin{tabular}{c c c}
\toprule
Parameter & Model & Prior \\
\midrule
$\nu$ & All & LogNormal(-2, 0.7) \\

$\lambda$ & All & Exponential(0.3) \\

$\rho$ & All & LogNormal(-3.00, 0.3) \\

$\kappa$ & All & Truncated-Normal(70, 80)\\

$\log R_1$ & All & Normal(0, 0.5) \\

$\sigma$ & RW1 & LogNormal(-0.6, 0.6)\\

$\sigma$ & OU & LogNormal(-2.6, 0.6)\\

$\theta$ & OU & Exp(1)\\

$\sigma$ & IBM & LogNormal(-0.5, 0.6)\\

$\sigma$ & RW2 & LogNormal(-2, 0.6)\\

$\sigma$ & HSGP & LogNormal(-0.6, 0.6)\\
$\ell$ & HSGP & Gamma(100, 20)\\
\bottomrule
\end{tabular}
\caption{Priors for hyperparameters in the simulation study.}
\label{table:priors}
\end{table}

Table \ref{table:priors} gives the priors used for the models fit to the simulated data. The number of basis functions $M$ and boundary scale $c$ for the HSGP were chosen using the procedure from \citep{Riutort-Mayol_etal_2022}. They were checked once for one simulation and then again for each county. Letting $l_{mean}$ be the mean length scale and $d$ be the length of the time series, we had
$c = \max\{4.5 \cdot l_{mean}/ (d/2), 1.2\}$
and $M = 3.45 \cdot c \cdot (d/2) /l_{mean}$.

\FloatBarrier

\subsection{MCMC Convergence for Models fit to Simulated Data}

\begin{figure}[htbp]
    \centering
    \includegraphics[width=0.9\textwidth]{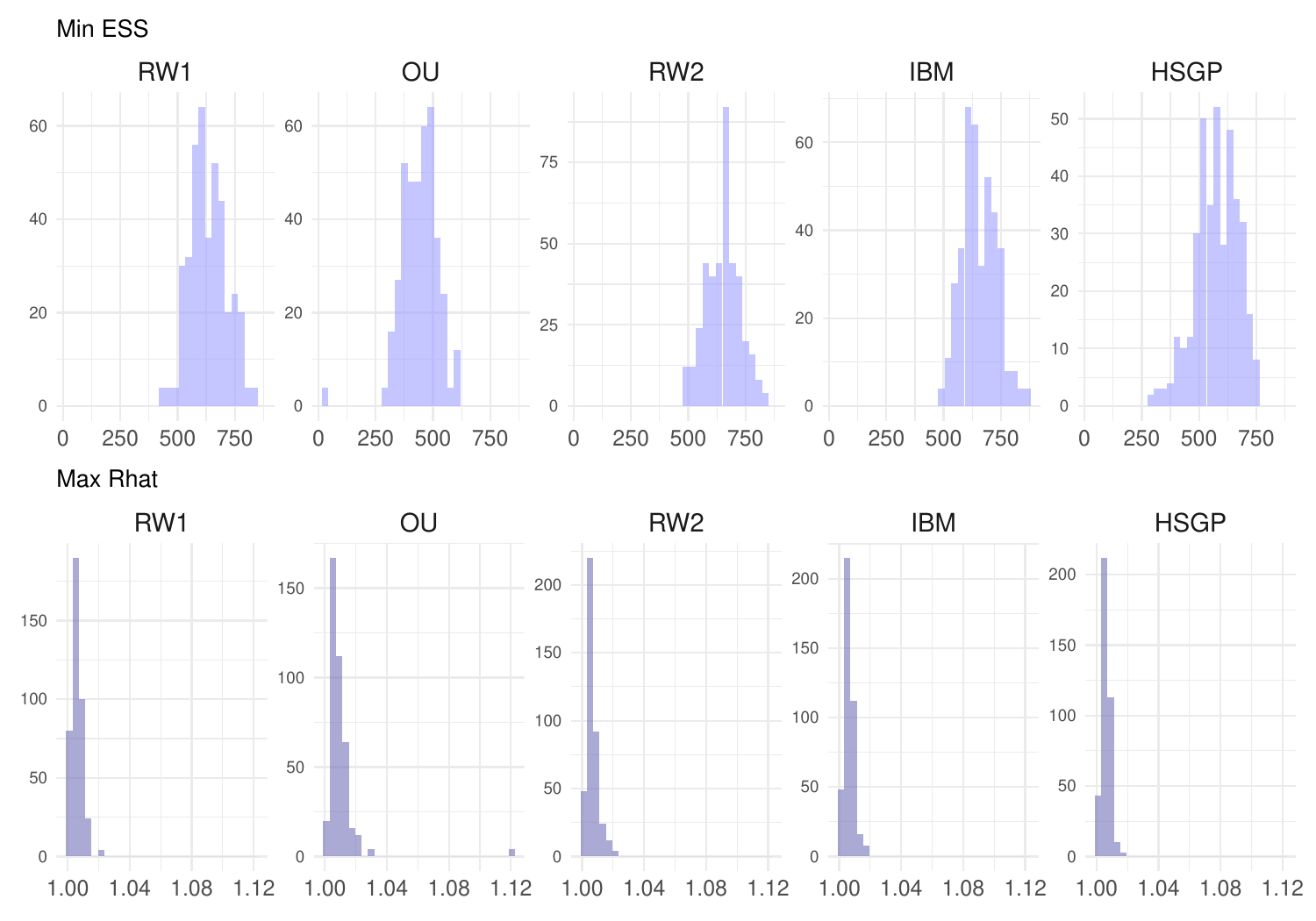}
    \caption{Histograms of minimum effective sample size (ESS) and maximum R-hat for the 100 simulated dataset fits.}
    \label{fig:mcmc_hists}
\end{figure}

We evaluated MCMC convergence primarily through traceplots, minimum ESS, and maximum Rhat. Convergence was deemed satisfactory if Rhat values were below 1.05, and an effective sample size greater than 250 was considered sufficient. Figure \ref{fig:mcmc_hists} summarises the minimum ESS and Rhat values for all MCMC runs in the simulation study, as traceplots are space-inefficient.

\FloatBarrier

\section{CPU Time Comparisons for Models fit to Simulated Data}
\label{sec:cputime}
\setcounter{table}{0}
\renewcommand{\thetable}{C\arabic{table}}
\setcounter{figure}{0}
\renewcommand{\thefigure}{C\arabic{figure}}
\begin{table}[htbp]
  \centering
  \begin{tabular}{lrrrrr}
  \toprule
    & Average Time & [Minimum Time, Maximum Time] \\
  \midrule
  RW1 & 3.08 & [1.65, 6.10] \\
  OU & 3.34 & [1.79, 5.08] \\
  RW2 & 4.06 & [2.15, 5.95] \\
  IBM & 6.41 & [2.05, 9.82] \\
  HSGP & 3.48 & [1.97, 5.10] \\
  EpiNow2 & 227.97 & [81.15, 276.90] \\
  \bottomrule
  \end{tabular}
  \caption{Mean (minimum, maximum) of CPU time (in minutes) for each model fit in the simulation study (100 models each). These models were fit on a time series of length 53 (weeks), with the exception of EpiNow2, which had to be fit to a time series of length 371 (see Appendix Section \ref{sec:epinow2} for details.)}
  \label{table:cpu_time}
\end{table}

\pagebreak

\section{Real Data Analysis Details}
\setcounter{table}{0}
\renewcommand{\thetable}{D\arabic{table}}
\setcounter{figure}{0}
\renewcommand{\thefigure}{D\arabic{figure}}

\subsection{Prior for $\kappa$ in the Analysis of San Francisco and Los Angeles Counties}
\label{sec:kappa}

The priors used in the real data analyses were mostly retained from the simulation study, with the exception of $\kappa$, the overdispersion of the negative binomial observation model. We follow the method of \citet{improving_rt}. Using \texttt{brms}, a Bayesian thin plate regression spline is fit to the time series of cases. The prior for $\kappa$ is then constructed using the 2.5\% and 97.5\% quantiles of posterior samples by minimising the squared difference between those quantiles and quantiles from a candidate truncated normal distribution, optimising the mean and standard deviation. Using synthetic data, the spline model can be fit to simulated data that are not directly used in the analysis. With real data, the spline can be fit to a location which is expected to be similar to the one for which the models will be run. For the purposes of this analysis, the spline model was fit to time series data from the same counties, but from a different time period for each.

\FloatBarrier

\subsection{Prior vs. Posterior Plots for Time-Invariant Parameters}

\subsubsection{Prior vs. Posterior Plots for Los Angeles County Analysis}

\begin{figure}[ht]
    \centering
    \includegraphics[width=0.88\textwidth]{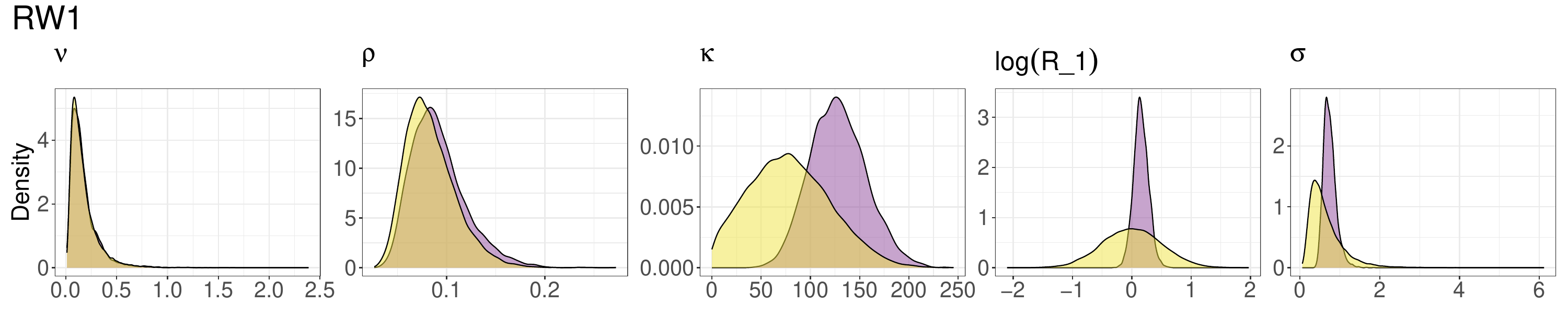}
    \includegraphics[width=0.88\textwidth]{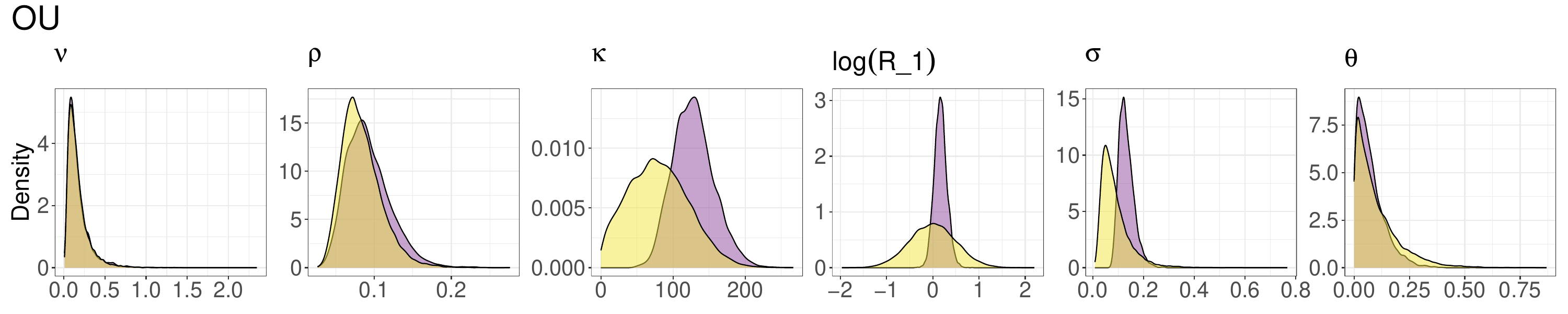}
    \includegraphics[width=0.88\textwidth]{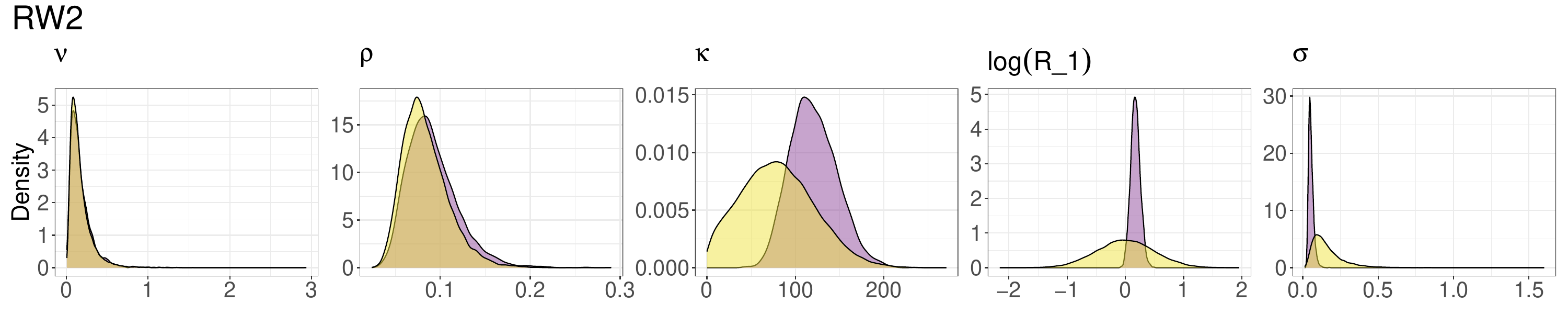}
    \includegraphics[width=0.88\textwidth]{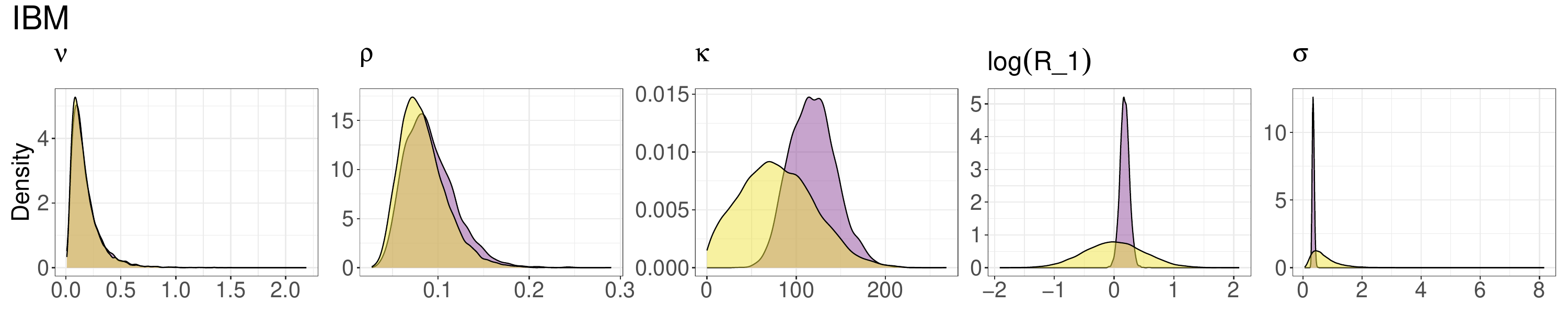}
    \includegraphics[width=0.88\textwidth]{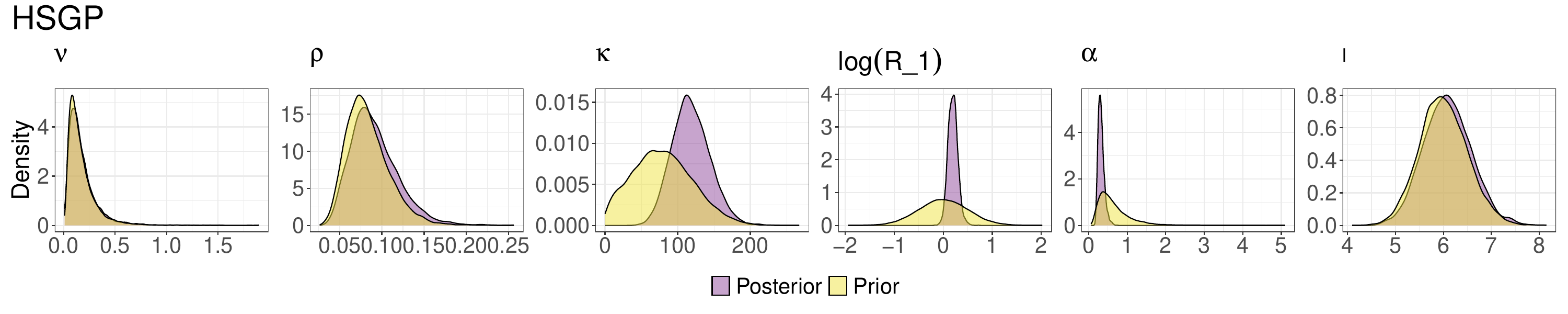}
    \caption{Prior vs. posterior plots for retrospective Los Angeles county analysis from Section \ref{sec:realdata}.}
    \label{fig:prior_v_posterior_la}
\end{figure}

\FloatBarrier

\subsubsection{Prior vs. Posterior Plots for San Francisco County Analysis}

\FloatBarrier

\begin{figure}[ht]
    \centering
    \includegraphics[width=0.88\textwidth]{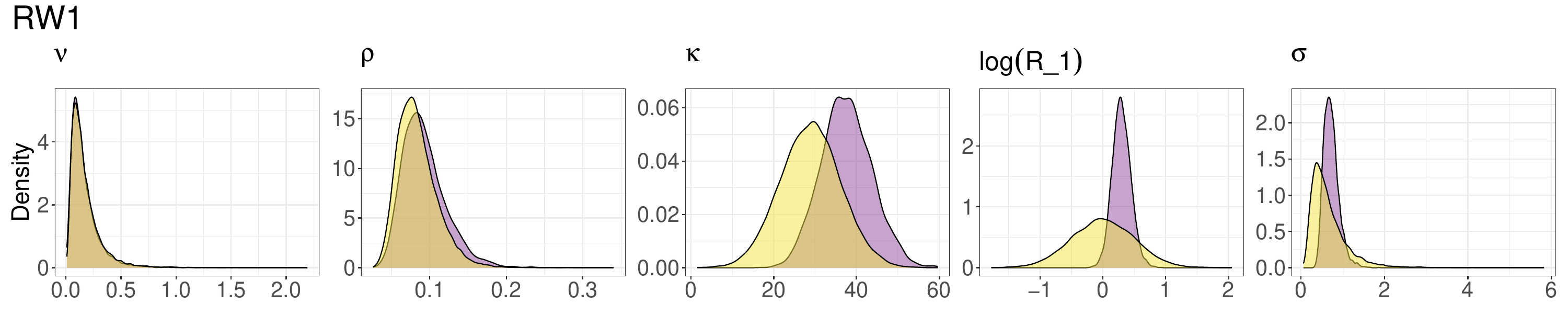}
    \includegraphics[width=0.88\textwidth]{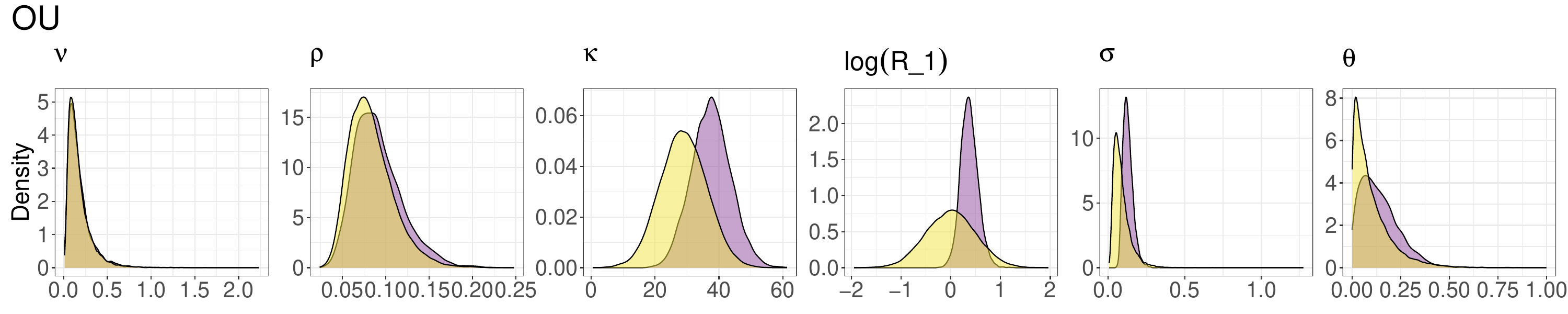}
    \includegraphics[width=0.88\textwidth]{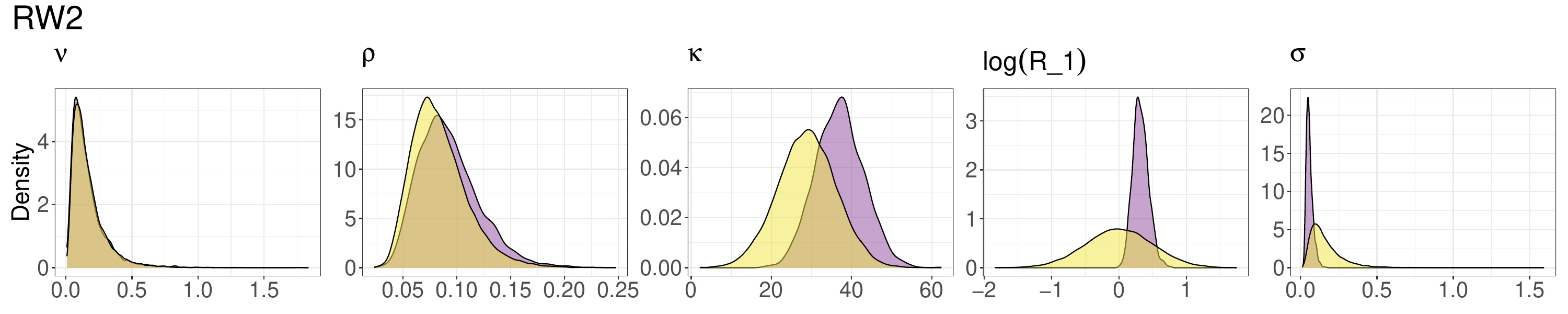}
    \includegraphics[width=0.88\textwidth]{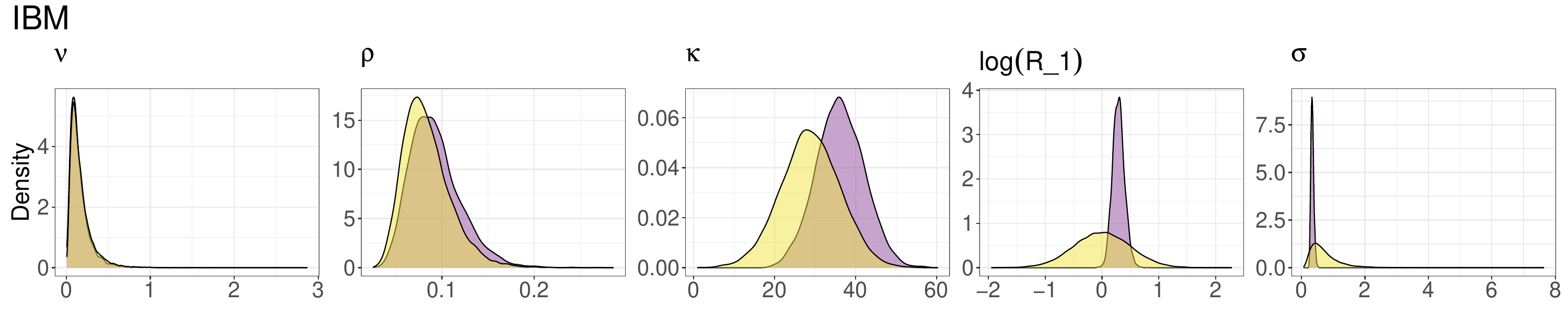}
    \includegraphics[width=0.88\textwidth]{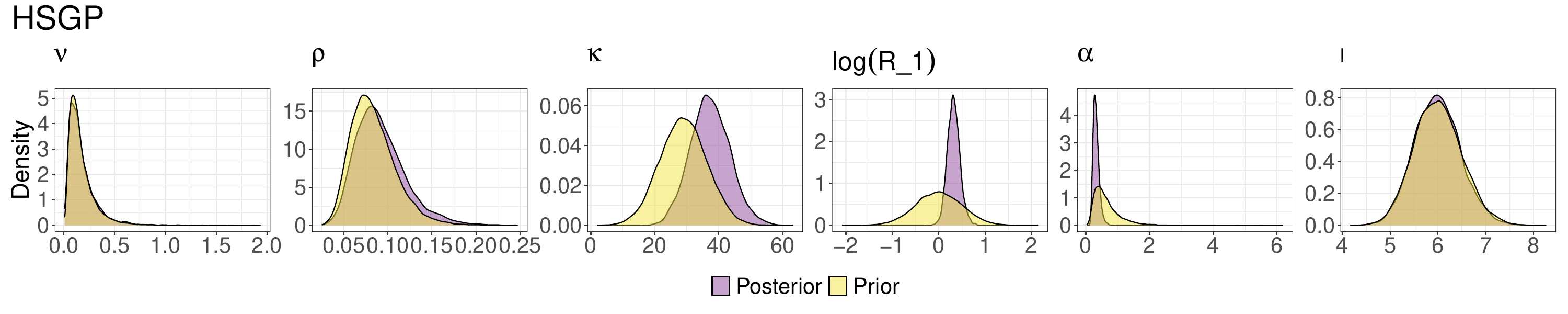}
    \caption{Prior vs. posterior plots for retrospective San Francisco county analysis from Section \ref{sec:realdata}.}
    \label{fig:prior_v_posterior_sf}
\end{figure}

\FloatBarrier

\subsection{MCMC Convergence}

\subsubsection{Traceplots for Log Posteriors}

\begin{center}
    \includegraphics[width=0.88\textwidth]{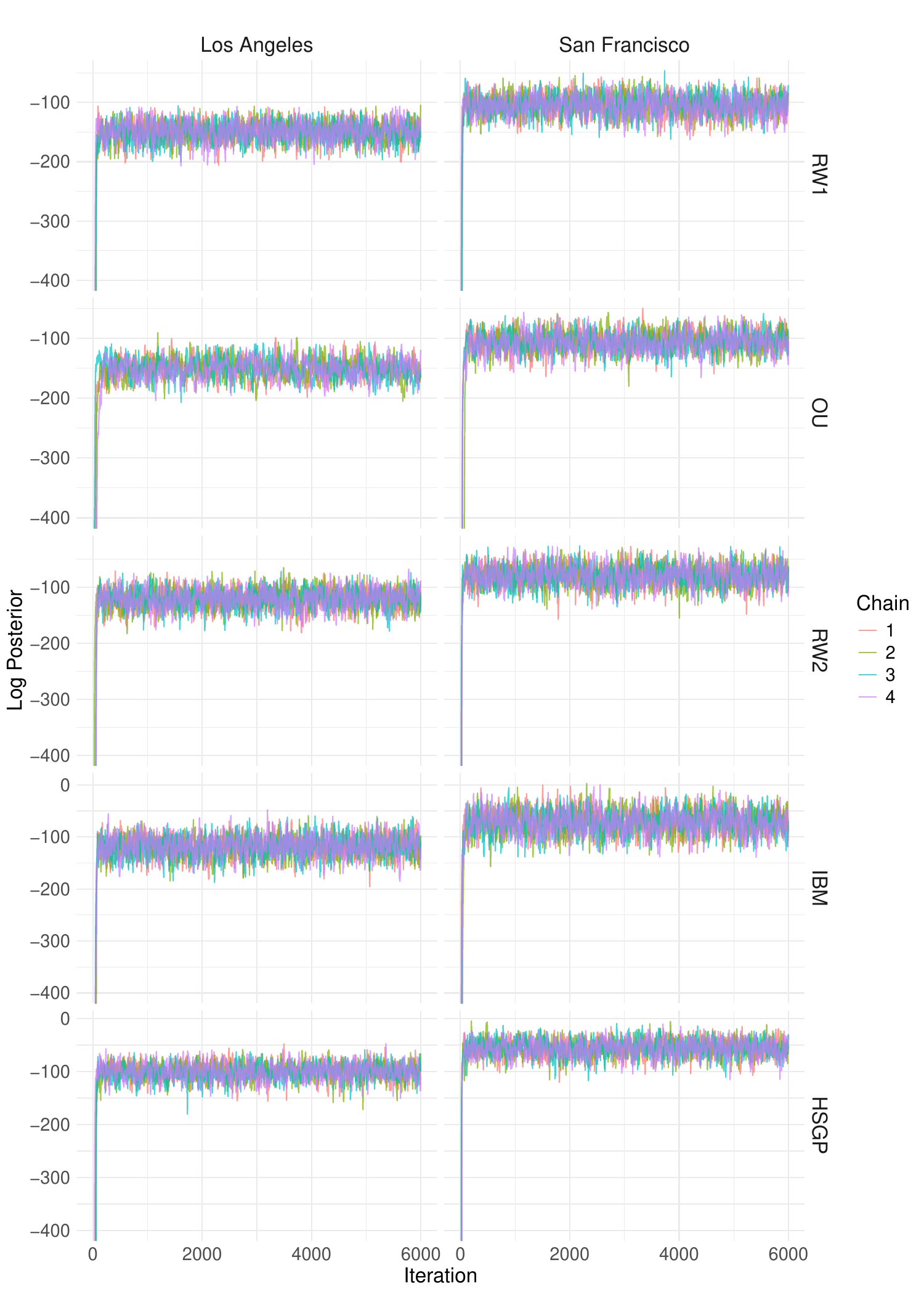}
    
    \captionof{figure}{Traceplots of the log posterior for the retrospective analyses of Los Angeles and San Francisco counties from Section \ref{sec:realdata}. Here we display only the log posterior traceplots due to space constraints, but all other parameters displayed similarly good mixing and fast convergence.}
    \label{fig:traceplots_realdata}
\end{center}

\FloatBarrier

\subsubsection{Table of MCMC Diagnostics for Los Angeles and San Francisco Analyses}

Table \ref{table:mcmc_diagnostics_county} summarises MCMC diagnostics for the real data analyses. All chains had fewer than 5 divergent transitions, and most had zero.

\begin{table}[ht]
  \centering
    \begin{tabular}{llrrrr}
    \toprule
    County & Prior & Divergent Transitions & Max Treedepth Hit & Max R-hat & Min ESS \\
    \midrule
    \multirow{5}{*}{Los Angeles} & RW1  & 0 & 3 & 1.007 & 913.277 \\
                        & OU   & 3 & 0 & 1.009 & 502.898 \\
                        & RW2  & 0 & 2 & 1.010 & 724.825 \\
                        & IBM  & 0 & 3 & 1.004 & 822.890 \\
                        & HSGP & 0 & 1 & 1.002 & 742.175 \\
    \midrule
    \multirow{5}{*}{San Francisco} & RW1  & 0 & 0 & 1.007 & 704.265 \\
                        & OU   & 1 & 34 & 1.013 & 572.900 \\
                        & RW2  & 0 & 0 & 1.008 & 722.933 \\
                        & IBM  & 0 & 0 & 1.003 & 506.352 \\
                        & HSGP & 0 & 0 & 1.003 & 711.496 \\
    \bottomrule
    \end{tabular}
    \caption{MCMC diagnostics for each prior and county.}
    \label{table:mcmc_diagnostics_county}
\end{table}

\subsection{Posterior Predictive Plots for Los Angeles and San Francisco Analyses}

\FloatBarrier

\begin{figure}[ht]
    \centering
    \includegraphics[width=\textwidth]{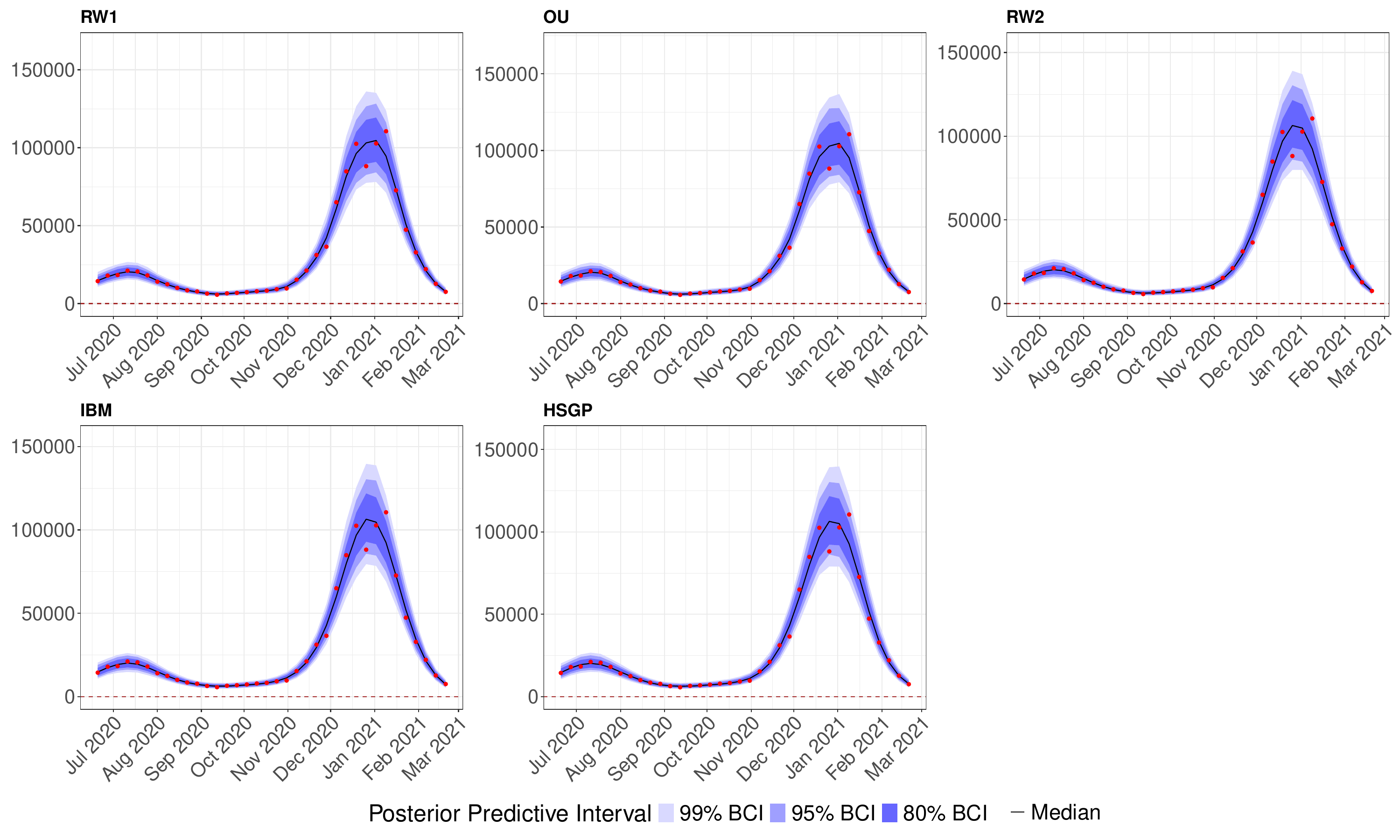}
    \caption{Posterior predictive plots for retrospective Los Angeles county analysis from Section \ref{sec:realdata}. Red points are observed case counts for Los Angeles county.}
    \label{fig:postpred_la}
\end{figure}

\begin{figure}[ht]
    \centering
    \includegraphics[width=\textwidth]{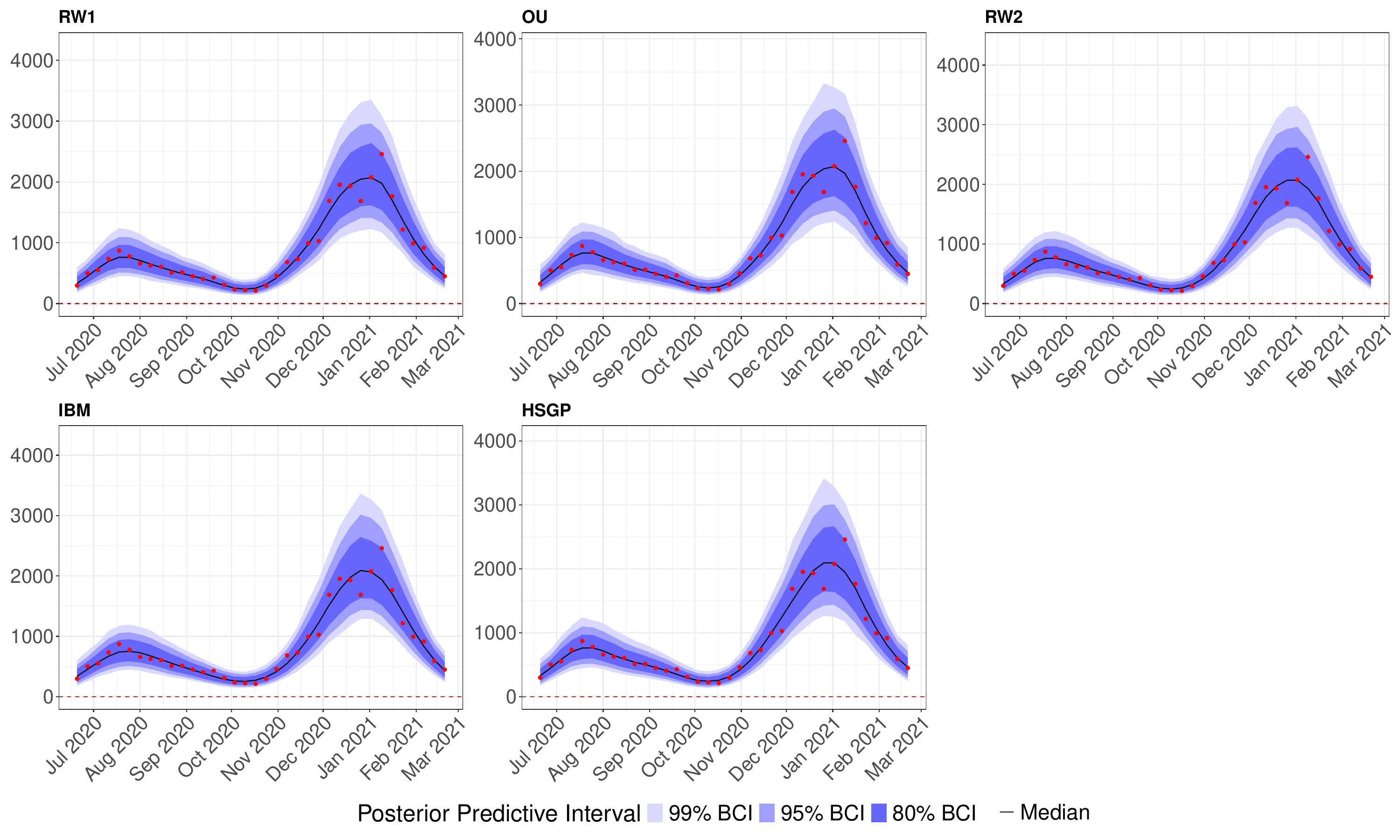}
    \caption{Posterior predictive plots for retrospective San Francisco county analysis from Section \ref{sec:realdata}. Red points are observed case counts for San Francisco county.}
    \label{fig:postpred_sf}
\end{figure}

\end{document}